\documentclass[12pt]{book}

\newcommand{\figWidth}{0.99\textwidth}

\newcounter{exampleCounter}

\usepackage[dvips]{color}
\usepackage{amsmath}
\usepackage{subfig}
\usepackage{graphicx} 
\usepackage[dvips]{color}
\usepackage{float}
\usepackage{graphicx}
\usepackage{tabularx}
\usepackage{pst-plot}
\usepackage{pstricks}

\usepackage{listings}
\usepackage{xcolor}

\usepackage{lipsum}
\usepackage{hyperref}

\title{Brain Gene Expression Analysis: a MATLAB toolbox for the analysis of brain-wide gene-expression data}
\author{\hspace{-5mm}Pascal Grange$^{1}$, Jason W. Bohland$ ^{3}$, Michael Hawrylycz$ ^{4}$, Partha P. Mitra$ ^{2}$\\
\footnotesize{\hspace{-5mm}$ ^{1}$ Xi'an Jiaotong-Liverpool University, Department of Mathematical Sciences,}\\
\footnotesize{\hspace{-5mm} 111 Ren'ai Rd, Suzhou, Jiangsu, China, 215123}\\
\footnotesize{\hspace{-5mm}$ ^{2}$ Boston University, College of Health \& Rehabilitation Sciences, Boston, MA 02215, United States}\\
\footnotesize{\hspace{-5mm}$ ^{3}$ Allen Institute for Brain Science,
Seattle, Washington 98103, United States}\\
\footnotesize{\hspace{-5mm}$ ^{4}$ Cold Spring Harbor Laboratory,
 Cold Spring Harbor, New York 11724, United States}\\
\normalsize{$^\ast$E-mail: {\ttfamily{pascal.grange@polytechnique.org}}}}

\begin{document}
\maketitle

\addtocounter{exampleCounter}{1}

\tableofcontents

\chapter{Preface}

The toolbox was first developed at Cold Spring Harbor Laboratory 
 and the Allen Institute for Brain Science as part of the research effort supported by the 
 NIH-NIDA Grant 1R21DA027644-01, {\emph{Co-expression networks of genes in the mouse and human brain}}. A presentation of the analysis techniques developed  for  the toolbox appeared in \cite{qbCoExpression}.\\

{\bf{Download instructions.}}  Data and code are available from {\hbox{\href{https://www.dropbox.com/sh/iwaiq9oxzd2jprl/AAB1oQjIgs7WHYrhnrB\_RRxea?dl=0}{Dropbox}}}.\\

$\bullet$ V1: presented at Neuroscience 2012, October 13-17, 2012 (New Orleans).\\

$\bullet$ V2: February 7, 2013. Corrected a bug in the first code snippet 
of Chapter 1 (the gene filter \lstinline$'allNoDup'$ has to be used,
 instead of \lstinline$'top75CorrNoDup'$, that was used in the initial 
 version of this manual). Thanks to Pablo Blinder (Tel Aviv University)
for raising the issue.\\

$\bullet$ V3: June 28, 2013. More typos corrected. Expanded 
section on fitting of brain regions by sets of genes. Thanks 
 to Andrew Martin (University of London) for correspondence.\\

$\bullet$ V4: July 28, 2014. Added two examples to Chapter 2, working 
 out regions corresponding to empty sets of voxels in the standard (hierarchical) 
 brain-wide annotation. References added.\\

$\bullet$ V5: July 3, 2015. Replaced download link. More typos corrected, thanks to Ran Bi (Xi'an Jiaotong-Liverpool
 University) for reporting them. In particular the symbol \lstinline{D}
  was used instead of \lstinline{E} in various places.\\

$\bullet$ V6: September 5, 2017. Fixed download link to avoid a cut-and-paste.\\


\chapter{The voxel-based Allen Atlas of the adult mouse brain}
\section{Presentation of the dataset}
\subsection{Co-registered \emph{in situ} hybridization data}

The adult mouse brain is partitioned into $V=49,742$ cubic voxels
of side 200 microns, to which {\emph{in situ}} 
hybridization data are registered \cite{ARA,AllenGenome} for thousands of genes.
For computational purposes, these gene-expression data can be arranged into 
 a voxel-by-gene matrix.\\

 For each voxel $v$, the {\it{expression energy}} of the gene $g$ is a
weighted sum of the greyscale-value intensities $I$ evaluated at the
pixels $p$ intersecting the voxel:
\begin{equation}
E(v,g) = \frac{\sum_{p\in v} M( p ) I(p)}{\sum_{p\in v} 1},
\label{ExpressionEnergy}
\end{equation}
 where $M( p )$ is a Boolean mask
 that equals $1$ if the gene is expressed at pixel
 $p$ and $0$ if it is not.\\

\subsection{Data matrices and gene filters}
\subsubsection{Coronal atlas}
 Some genes in the Allen atlas of the adult mouse brain gave rise
 to an ISH experiment and coronal sectioning of an entire brain.
 The resulting data constitute the {\emph{coronal atlas}}.
 The coronal atlas contains brain-wide data for $G_{\mathrm{all}} = 4,104$ genes (this is a subset 
 of the genes for which ISH sagittal sectioning took place for a hemisphere
  after ISH, see next section for documentation on the sagittal atlas).
 The corresponding voxel-by-gene data matrix has size $V=49,742$ by
 $G_{\mathrm{all}}$, and is contained in the file
 \lstinline$ExpEnergy.mat$. The list of genes arranged in the same order as the columns
 of the data matrix are obtained by using the function \lstinline$get_genes.m$.\\

\lstset{language=Matlab,basicstyle=\ttfamily,commentstyle=\bf,
captionpos=b,tabsize=7,frame=lines,numbers=left,numbersep=0.05pt,
numberstyle=\tiny,numbersep=0.025pt,
breaklines=false,showstringspaces=false,
basicstyle=\ttfamily\footnotesize,emph={label}}   
\begin{lstlisting}
load( 'ExpEnergy.mat' );
% the g-th column of voxel-by-gene matrix E corresponds
%  to the gene geneNamesAll( g )
geneNamesAll = get_genes( Ref, 'allNoDup', 'allen' );
% Entrez Ids are arranged in the same order
% as gene names (unresolved Entrez ids are treated as zero)
geneEntrezIdsAll = get_genes( Ref, 'allNoDup', 'entrez' );
\end{lstlisting}

The matrix in \lstinline$ExpEnergyTop75Percent.mat$ consists 
of the 3,041 columns of the matrix defined in Equation \ref{ExpressionEnergy} 
that are best correlated with 
the corresponding genes in the sagittal atlas. 
 The names and Entrez ids of the genes\footnote{The differences between the two 
 code snippets are the data matrix loaded, and the filter used
 as teh second argument of the function \lstinline$get_genes$. One can check
 that after executing any of the snippets, the number of gene names
 equal the number of columns of the data matrix.} are obtained as follows:\\
\begin{lstlisting}
load( 'ExpEnergytop75Percent.mat' );
% the g-th column of voxel-by-gene matrix E
% corresponds to the gene genesAllen( g )
genesAllen = get_genes( Ref.Coronal, 'top75corrNoDup', 'allen' );
% Entrez Ids are arranged in the same order
%  as gene names (unresolved Entrez ids are treated as zero)
genesEntrez = get_genes( Ref.Coronal, 'top75corrNoDup', 'entrez' );  
\end{lstlisting}
It should be noted that the Entrez ids that are not resolved are represented by
 zeroes in \lstinline$geneEntrezIds$, so Entrez ids should be resolved
 {\emph{in fine}} rather than used during computations.\\

The start-up file \lstinline$mouse_start_up.m$ loads
the Allen Reference Atlas stored in the 
 file \lstinline$refAtlas.mat$ and the data matrix 
 \lstinline$ExpEnergytop75percent.mat$,
 which can be used through the structure \lstinline$Ref$.

$\bullet$ {\bf{Example \arabic{exampleCounter}. Systems of brain 
 annotation.\addtocounter{exampleCounter}{1}}}
The variable \lstinline$cor=Ref.Coronal$ contains the coronal atlas, and
 it can be used to check some of the contents of Table \ref{annotationSystems}.
 a detailed description of the fields in the structure
 \lstinline$Ref.Coronal.Annotations$ can be found in the next section.\\
\begin{lstlisting}
mouse_start_up;
cor = Ref.Coronal;
ann = cor.Annotations;
display( ann );
\end{lstlisting}

\begin{table}
\centering 
\begin{tabular}{|m{0.12\textwidth}|m{0.2\textwidth}|m{0.11\textwidth}|m{0.2\textwidth}|m{0.2\textwidth}|}
\hline
\textbf{Identifier index}& \textbf{Identifier (name of annotation)}&\textbf{Number of regions}
&\textbf{Hierarchical?}& \textbf{Number of annotated voxels}\\ \hline
1 & {\ttfamily{’standard’}} & 209 & Yes &  49,742\\ \hline
2 & {\ttfamily{’cortex’}} & 40 &  Yes & 11,862\\ \hline
3 & {\ttfamily{’standard+cortex’}} & 242  & Yes &  49,742\\ \hline
4 & {\ttfamily{’fine’}} & 94 &  No  & 22,882\\ \hline 
5 & {\ttfamily{’big12’}} & 13 &  No &  25,155 \\ \hline
6 & {\ttfamily{’cortexLayers’}} & 8 & No & 11,862 \\ \hline
\end{tabular}
\caption{Systems of annotations of the adult mouse brain in the digital version
of the Allen Reference Atlas, at a resolution of 200 microns.}
\label{annotationSystems}
\end{table}



\section{From matrices to volumes}
\subsection{The Allen Reference Atlas (ARA)}
 The ARA comes in six different versions, described
 in Table \ref{annotationSystems}.
 Each of these versions corresponds to an annotation 
 of the three-dimensional grid by digital ids. Each of these digital ids
 corresponds to a brain region. The correspondence 
 between ids and names of brain regions can be resolved using 
 the fields \lstinline$Ref.Coronal.ids$ and \lstinline$Ref.Coronal.labels$.\\

$\bullet$ {\bf{Example \arabic{exampleCounter}. From brain regions
 to digital ids in the ARA.\addtocounter{exampleCounter}{1}}} 
Consider the \lstinline$’fine’$ annotation (identifier index 4 in Table 1) and let us
work out a three-dimensional grid with ones at the voxels corresponding to the caudoputamen in this annotation, and zeroes everywhere else. We can use this grid to compute the number of voxels in the caudoputamen.\\

\begin{lstlisting}
% the three-dimensional grid containing the fine annotation of the brain
annotFine = get_annotation( cor, 'fine' );
% the names of brain regions
labels = ann.labels{ 4 };
% the numerical ids of the regions
ids = ann.ids{ 4 };
% where is the caudoputamen in the list? There may be spaces 
% in the labels, skip them
labelsNoSpace = regexprep( labels, '\W', '' );
caudouputamenIndex = find( strcmp( labelsNoSpace, 'Caudoputamen' ) == 1 );
caudouputamenId = ids( caudouputamenIndex );
% put zeros at all the voxels that do not belong to caudoputamen
volCaudoputamen = annotFine;
volCaudoputamen( volCaudoputamen ~= caudouputamenId ) = 0;
volCaudoputamen( volCaudoputamen == caudouputamenId ) = 1;
% count the voxels in caudoputamen
numVoxInCaudoputamen = sum( sum( sum( volCaudoputamen ) ) );
display( numVoxInCaudoputamen )
\end{lstlisting}
The value of the variable \lstinline$numVoxInCaudoputamen$ computed in 
the above code snippet should be 1248.\\

\subsection{From gene-expression vectors to volumes}
At a resolution of 200 microns, the Allen Reference Atlas (ARA)
is embedded in a three-dimensional grid of size $67\times 41\times 58$.
 Out of the $V_{\mathrm{tot}}=67\times 41\times58$ in the grid, $V=49,742$ are in 
 in the brain according to the ARA.\\

The brain voxels can be mapped to a three-dimensional grid using the 
function\\
\lstinline$make_volume_from_labels.m$
 and a specified voxel filter, corresponding to one of the 
versions of the ARA (as per Table \ref{annotationSystems}).\\

$\bullet$ {\bf{Example \arabic{exampleCounter}. Whole-brain
    filter.\addtocounter{exampleCounter}{1}}} Let us take the whole
first column of the data matrix and map it to a three-dimensional
grid, using the whole-brain filter, corresponding to the 'standard'
annotation:\\
\begin{lstlisting}
wholeBrainFilter = Ref.Coronal.Annotations.Filter{ 1 };
display( wholeBrainFilter );
brainFilter = get_voxel_filter( cor, wholeBrainFilter );
% a column vector with 49,742 elements
col1 = E( :, 1 );
display( size( col1 ) ) 
% map this column vector to a volume
vol1 = make_volume_from_labels( col1, brainFilter );
display( size( vol1) );
\end{lstlisting}

 Some of the annotations do not extend to the whole brain, as can be
 seen from Table \ref{annotationSystems}.  when comparing columns of
 the matrix of gene-expression energies to regions in the ARA, it is
 important to restrict the matrix to the rows that correspond to
 annotated voxels. For each system of annotation, the field
 \lstinline$Ref.Coronal.Annotations.Filter$ is the list of voxels 
 in a $67\times 41\times 58$ grid ({\emph{not}} a list of row indices 
in the matrix of expression energies), that are annotated. 
The example below shows how to recover
 the list of row indices in the gene-expression matrix corresponding to 
 a given filter.\\

$\bullet$ {\bf{Example \arabic{exampleCounter}.
 More filters.\addtocounter{exampleCounter}{1}}}
 For instance, one can check that the following two snippets
 produce the same matrix \lstinline$EFiltered$, consisting of the 
 rows of the full gene-expression matrix corresponding to 
 voxels in the \lstinline$fine$ annotation:\\
\begin{lstlisting}
 % Compute the set of rows using a volume of 
 %indices and the filter 
 cor = Ref.Coronal;
 identifierIndex = 4;
 brainFilter = get_voxel_filter( cor, 'brainVox' );
 numVox = numel( brainFilter );
 % label the voxels by integers
 indsBrainVoxels = 1 : numVox;
 % arrange the integers in a volume
 indsVol = make_volume_from_labels( indsBrainVoxels, brainFilter );
 filter = get_voxel_filter( cor, ann.filter{ identifierIndex } );
 % restrict the volume to the voxels that are in the filter
 indsFiltered = indsVol( filter );
 EFiltered = E( indsFiltered, : );}}
\end{lstlisting}

\begin{lstlisting}
 %apply the filter to each column of the data matrix
 cor = Ref.Coronal;
 identifierIndex = 4;
 brainFilter = get_voxel_filter( cor, 'brainVox' );
 filter = get_voxel_filter( cor, ann.filter{ identifierIndex } );
 numGenes = size( E, 2 );
 % restrict each colum of the data matrix to voxels
 %  that are in the filter
 for gg = 1 : numGenes
    geneVol = make_volume_from_labels( E( :, gg ), brainFilter );
    EFiltered( :, gg ) = geneVol( filter );
 end
\end{lstlisting}

Note also that the voxel filter can be recomputed from the 
three-dimensional annotation and the brain-wide filter:\\
\begin{lstlisting}
filterStored = get_voxel_filter( cor, 'leftVox' );
brainFilter = get_voxel_filter( cor, 'brainVox' );
annot = get_annotation( cor, ann.identifier{ identifierIndex } );
annotArray = annot( brainFilter );
% the annotated voxels are filled with non-zero numerical ids
annotatedIndices = find( annotArray~= 0 );
filterComputed = brainFilter( annotatedIndices );
\end{lstlisting}

\subsection{Visualization}
\subsubsection{Maximal-intensity projection}
 Given a three-dimensional grid of size $67\times 41 \times 58$, 
its maximal-intensity projection onto sagittal, coronal and axial planes can be plotted
 using the function\\
 \lstinline$plot_intensity_projections.m$\\

\begin{figure}
\centering
\includegraphics[width=\figWidth,keepaspectratio]{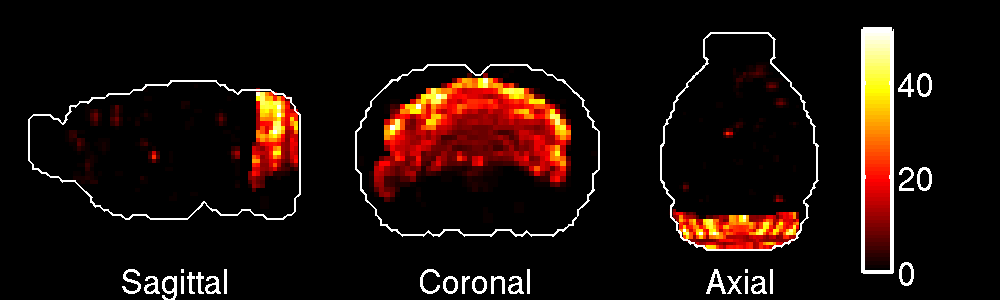}
\caption{Maximal-intensity projections of the expression energy of {\emph{Gabra6}}.}
\label{Gabra6Projection}
\end{figure}

$\bullet$ {\bf{Example \arabic{exampleCounter}. Maximal-intensity projections of
    the expression energy of a gene.\addtocounter{exampleCounter}{1}}}
Let us plot maximal-intensity projections of the expression energy of
{\emph{Gabra6}}. The abovefollowing code snippet should reproduce
Figure \ref{Gabra6Projection}.\\
\begin{lstlisting}
% find out to which column of the data matrix it corresponds
indexGabra6 = find( strcmp( genesAllen, 'Gabra6' ) == 1 );
columnGabra6 = E( :, indexGabra6 );
% map the column to a volume
volGabra6 = make_volume_from_labels( columnGabra6, brainFilter );
plot_intensity_projections( volGabra6 );
\end{lstlisting}

\begin{figure}
\centering
\includegraphics[width=\figWidth,keepaspectratio]{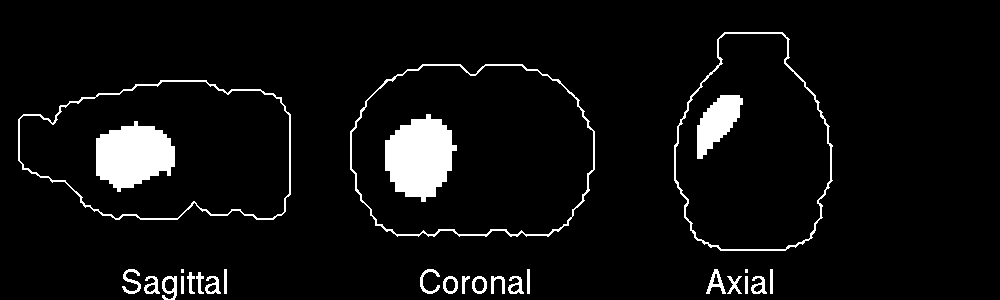}
\caption{Maximal-intensity projections of the characteristic function of the caudoputamen in
the \lstinline$fine$ annotation.}
\label{projectionCaudoputamenFine}
\end{figure}

$\bullet$ {\bf{Example \arabic{exampleCounter}. Maximal-intensity projections of brain regions\addtocounter{exampleCounter}{1}}} We can use the function \lstinline$plot_intensity_projections.m$ to visualize brain regions in the atlas. For instance, we can use the 
 grid \lstinline$volCaudoputamen$ computed in one of the examples 
 above and reproduce Figure \ref{projectionCaudoputamenFine}.\\
\begin{lstlisting}
% the three-dimensional grid containing the fine annotation of the brain
annotFine = get_annotation( cor, 'fine' );
labels = ann.labels{ 4 };
ids = ann.ids{ 4 };
labelsNoSpace = regexprep( labels, '\W', '' );
caudouputamenIndex = find( strcmp( labelsNoSpace, 'Caudoputamen' ) == 1 );
caudouputamenId = ids( caudouputamenIndex );
volCaudoputamen = annotFine;
volCaudoputamen( volCaudoputamen ~= caudouputamenId ) = 0;
volCaudoputamen( volCaudoputamen == caudouputamenId ) = 1;
plot_intensity_projections( volCaudoputamen );
\end{lstlisting}

\begin{figure}
\centering
\includegraphics[width=\figWidth,keepaspectratio]{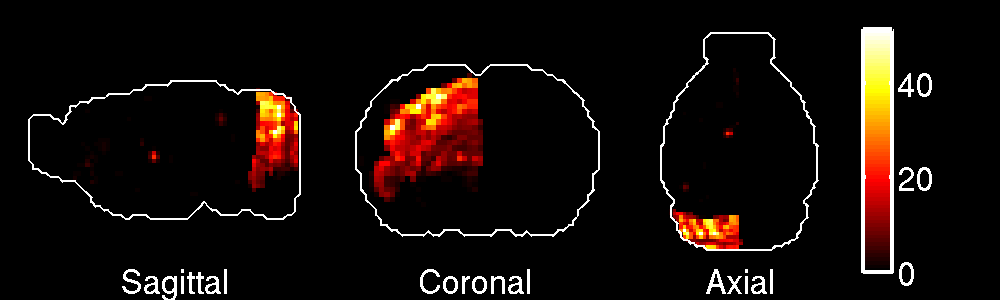}
\caption{Maximal-intensity projections of the expression energy of {\emph{Gabra6}}.}
\label{projectionGabra6Fine}
\end{figure}

\begin{figure}
\centering
\includegraphics[width=\figWidth,keepaspectratio]{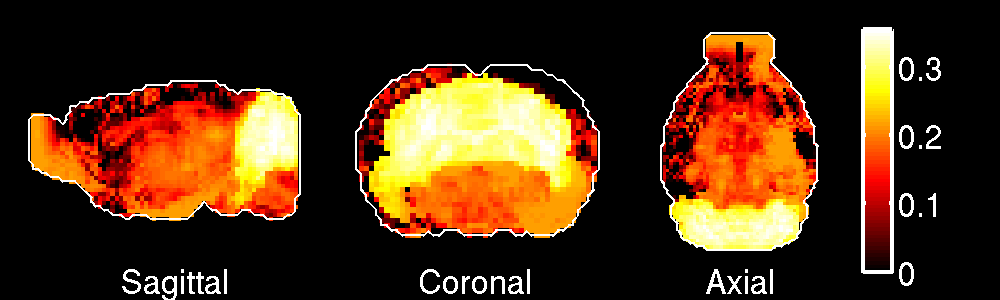}
\caption{Maximal-intensity projections of correlations between the voxels of the coronal  Allen Atlas (G=3,041 genes) and granule cells.}
\label{Gabra6example}
\end{figure}

{\bf{Example \arabic{exampleCounter}. \addtocounter{exampleCounter}{1} The 
\lstinline$fineAnatomy$ filter.}} It is manifest 
from the projection of the characteristic function of the caudoputamen
 in the \lstinline$fine$ annotation (Figure \ref{projectionCaudoputamenFine})
 that the \lstinline$fine$ annotation contains only the left hemisphere of the brain.
 We can confirm this by applying the voxel filter of the 
  \lstinline$fine$ annotation (called \lstinline$fineAnatomy$, which is the
 value of \lstinline$Ref.Coronal.Annotations.filter{4}$) 
to the gene-expression vector of {\emph{Gabra6}}. The following 
snippet should reproduce Figure \ref{projectionGabra6Fine}, which is equivalent 
 to Figure \ref{Gabra6example}, with all the voxels that are not in the
 \lstinline$fineAnatomy$ filter filled with zeros.\\

\begin{lstlisting}
% take the brain-wide expression of Gabra6
indexGabra6 = find( strcmp( genesAllen, 'Gabra6' ) == 1 );
columnGabra6 = E( :, indexGabra6 );
volGabra6 = make_volume_from_labels( columnGabra6, brainFilter );
% consider the voxel filter of the 'fine' annotation
fineFilter = get_voxel_filter( cor, 'fineAnatomy' );
% take the expression values at the voxels in the filter, and arrange them
% in a column vector (one can check that
% dataGabra6Filtered has the same number of
% elements as fineFilter
colGabra6Filtered = volGabra6( fineFilter );
% map these values to a three-dimensional grid, using fineFilter
volGabra6Filtered = make_volume_from_labels( colGabra6Filtered,...
  fineFilter );
plot_intensity_projections( volGabra6Filtered );
\end{lstlisting}

\begin{figure}
\centering
\includegraphics[width=\figWidth,keepaspectratio]{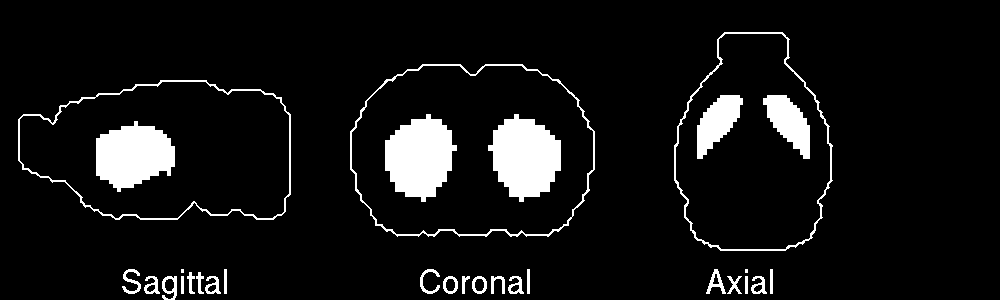}
\caption{Maximal-intensity projections of the caudoputamen in
the \lstinline$standard$ annotation.}
\label{projectionCaudoputamenStandard}
\end{figure}

$\bullet$ {\bf{Example
    \arabic{exampleCounter}. \addtocounter{exampleCounter}{1} Brain-wide
    versus left hemisphere.}} As can be seen on Table
\ref{annotationSystems}, the \lstinline$standard$ annotation
contains all the voxels in the brain (49,742 of them). We can compute
the characteristic function of the caudoputamen in this annotation,
and check that the left half of the caudoputamen coincides
 with Figure \ref{projectionCaudoputamenFine}, and that the number of voxels
is twice the number computed in the \lstinline$fine$ annotation.
 The code snipped below should reproduce 
Figure \ref{projectionCaudoputamenStandard}.\\

\begin{lstlisting}
identifierIndex = 1;
annotStandard = get_annotation( cor, ann.identifier{ identifierIndex });
ann = cor.Annotations;
labels = ann.labels{ identifierIndex };
% their numerical ids
ids = ann.ids{ identifierIndex };
% where is the caudoputamen in the list?
labelsNoSpace = regexprep( labels, '\W', '' ) ;
caudouputamenIndex = find( strcmp( labelsNoSpace, 'Caudoputamen' ) == 1 );
caudouputamenId = ids( caudouputamenIndex );
% put zeros at all the voxels that do not belong to caudoputamen
volCaudoputamen = annotStandard;
volCaudoputamen( volCaudoputamen ~= caudouputamenId ) = 0;
volCaudoputamen( volCaudoputamen == caudouputamenId ) = 1;
% count the voxels in caudoputamen
numVoxInCaudoputamenStandard = sum( sum( sum( volCaudoputamen ) ) );
display( numVoxInCaudoputamenStandard );
% reproduce Figure \ref{projectionCaudoputamenStandard}
plot_intensity_projections( volCaudoputamen );
\end{lstlisting}

\subsubsection{Sections}
The function \lstinline$flip_through_sections.m$
allows to go through the sections of a ($67\times 41\times 58$)
volume, of a kind specified in the options. It pauses between sections.
 The duration of the pause is one second by default, it can
 be adjusted using the field \lstinline$secondsOfPause$ of the options.
 If the value of \lstinline$secondsOfPause$ is negative,
 the user will have to press a key to display the next section.\\ 

\begin{figure}
\centering
\includegraphics[width=\figWidth,keepaspectratio]{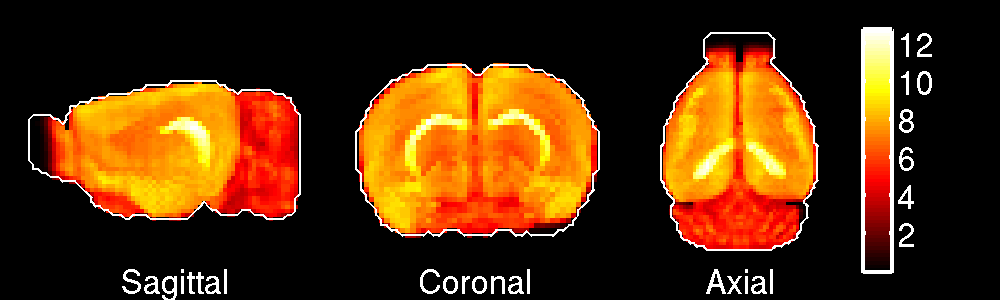}
\caption{Maximal-intensity projections of the average gene-expression $\overline{E}$
defined in Equation \ref{avgExprDef}.}
\label{avgExpressionProj}
\end{figure}

$\bullet$ {\bf{Example
    \arabic{exampleCounter}.\addtocounter{exampleCounter}{1}}}
Sections of the average across all genes in the dataset.
The following code allows to visualize the coronal, sagittal 
 and axial sections of the average expression $\overline{E}$  across
 all genes in the data matrix, defined in Equation \ref{avgExprDef}:
\begin{equation}
\overline{E}( v ) = \frac{1}{G}\sum_{g=1}^GE(v,g).
\label{avgExprDef}
\end{equation} 
It also reproduces the maximal-intensity projections of $\overline{E}$,
shown in Figure \ref{avgExpressionProj}.

\begin{lstlisting}
% the average gene-expression vector;
numGenes = size( E, 2 );
avgExpressionVector = sum( E, 2 ) / numGenes;
% map it to a volume
brainFilter = get_voxel_filter( Ref.Coronal, 'brainVox' );
avgExpressionVol = make_volume_from_labels( avgExpressionVector, brainFilter );
% visualize projections of this volume
plot_intensity_projections( avgExpressionVol );
hold off;
%flip through coronal sections, pause one second between sections
optionsCoronal = struct( 'sectionStyle', 'coronal', 'secondsOfPause', 1 );
flipThroughSections = flip_through_sections( avgExpressionVol, optionsCoronal );
close all;
%flip through sagittal sections
optionsSagittal = struct( 'sectionStyle', 'sagittal', 'secondsOfPause', 1 );
flipThroughSections = flip_through_sections( avgExpressionVol, optionsSagittal );
close all;
%flip through axial sections
optionsAxial = struct( 'sectionStyle', 'axial', 'secondsOfPause', 1 );
flipThroughSections = flip_through_sections( avgExpressionVol, optionsAxial );
close all;
\end{lstlisting}

\section{Relation between the various annotations in the ARA}

 The field \lstinline$Ref.Coronal.Annotations.parentIdx$ in the reference data
structure gives the indices (not the numerical id) of the parents of
the regions in a given annotation, arranged in the same order as these
regions. If a region does not have a parent in the considered system
of annotation, the corresponding entry in \lstinline$parentIdx$ is
zero.\\ 
NB: the list of regions in the 'cortex' annotation is not
closed under the operation of taking the parent of a region in a
hierarchy. The field \lstinline$parentSymbols$ has to be used instead
of parentIdx to work out the hierarchy.\\

$\bullet$ {\bf{Example \arabic{exampleCounter}. \addtocounter{exampleCounter}{1} Hierarchical 
and non-hierarchical annotations.}} Let us check that the \lstinline$big12$ and 
\lstinline$fine$ 
annotations are non-hierarchical, and investigate parents and descendants 
 of the caudoputamen and of the cerebellum in the \lstinline$standard$ annotation. 
\begin{lstlisting}
ann = Ref.Coronal.Annotations; 
% the parent indices in the 'big12' annotation are zeroes 
parentIdxBig12 = ann.parentIdx{ 5 } 
% check that the 'big12' annotation is non-hierarchical
Big12IsNonHierarchical = isempty( setdiff( unique( parentIdxBig12 ),...
 [ 0 ] ) ) 
% same goes for the 'fine' annotation 
parentIdxFine = ann.parentIdx{ 4 } 
% check that the 'fine' annotation is non-hierarchical
fineIsNonHierarchical = isempty( setdiff( unique( parentIdxBig12 ),...
 [ 0 ] ) ) 
% work out the names of the parents of a region in the standard 
% annotation 
labelsStandard = ann.labels{ 1 }; 
labelsStandardNoSpace = regexprep( labelsStandard, '\W', '' ) ; 
indexCaudouputamen = find( strcmp( labelsStandardNoSpace,...
                      'Caudoputamen' ) == 1 ); 
parentIndicesStandard = ann.parentIdx{ 1 }; 
parentIndexCaudoputamen = parentIndicesStandard( indexCaudouputamen ); 
display( labelsStandard( parentIndexCaudoputamen ) ) 
% and the names of the descendants 
caudoputamenDescendantIndices =
find( parentIndicesStandard == indexCaudouputamen ); 
% Is caudoputamen a leaf in the tree aunderlying the annotation? 
caudoputamenIsALeaf = isempty( caudoputamenDescendantIndices );
display( caudoputamenIsALeaf );
% What are the subregions of the cerebellum? 
cerebellumIndex = find( strcmp( labelsStandardNoSpace, 'Cerebellum' ) == 1 ) 
cerebellumDescendantIndices = find( parentIndicesStandard == cerebellumIndex );
cerebellumDescendants = labels( cerebellumDescendantIndices ); 
display( cerebellumDescendants ); 
\end{lstlisting}

$\bullet$ {\bf{Example \arabic{exampleCounter}. \addtocounter{exampleCounter}{1} Working out empty regions in a hierarchical annotation.}} Let us 
 work out which regions (if any) correspond to an empty set of voxels in the \lstinline$standard$ annotation.\\

\begin{lstlisting}
% Consider the 'standard' annotation, pull out numerical 
% ids and labels of brain regions
identifierIndex = 1;
cor = Ref.Coronal;
ann = cor.Annotations;
identifier = ann.identifier{ identifierIndex };
ids = ann.ids{ identifierIndex };
labels = ann.labels{ identifierIndex };

% Pull out the the annotated voxelized brain volume 
annot = get_annotation( cor, identifier );

% what are the brain regions that correspond to no voxel
% in the voxelized brain volume?
unresolvedIndices = [];
nRegions = numel( ids );
for rr = 1 : nRegions
   idRegion = ids( rr );
   annotCopy = annot;
   annotCopy( annotCopy ~= idRegion ) = 0;
   annotCopy( annotCopy == idRegion ) = 1;
   nVoxelsInRegion = sum( annotCopy( : ) );
   if nVoxelsInRegion == 0
     unresolvedIndices = [ unresolvedIndices, rr ];
     nEmptyRegions = numel( unresolvedIndices );
     unresolvedRegions{ nEmptyRegions } = char( labels( rr ) );
   end
end   
nUnresolvedRegions = numel( unresolvedRegions );
\end{lstlisting}
One can check that the variable \lstinline$nUnresolvedRegions$ at the end of the above code snippet equals $7$. 
  Let us work out which of these regions are terminal nodes (leaves of the hierarchy)
\begin{lstlisting}
% How are the unresolved regions situated in the hierarchy?
parentIndices = ann.parentIdx{ identifierIndex };
for uu = 1 : nUnresolvedRegions
    uu
    unresolvedIndex = unresolvedIndices( uu );
    display( labels{unresolvedIndex} );
    % which are the descendents?
    find( parentIndices == unresolvedIndex );
    %display( unresolvedRegions{ uu } );
    descendantIndices{ uu } = find( parentIndices == unresolvedIndex );
    if ~isempty( descendantIndices{ uu } )        
        descendantRegionsLoc = labels( descendantIndices{ uu } );
    else
       descendantRegionsLoc = '';
       display( [ labels{unresolvedIndex},  ' does not have any subregion' ] );
       pause;
    end
    descendantIndices{ uu } = descendantRegionsLoc;
    display( descendantRegionsLoc );
end    
\end{lstlisting}
 Only two of the $7$ regions corresponding to an empty  set of voxels 
  are therefore leaves of the hierarchical tree. We can conclude that they are too small to be represented by 
   voxels at a spatial resolution of 200 microns. The other 5 regions have descendants
    in the hierarchy, but they can only be resolved by taking the reunion 
     of the voxels belonging to their descendants (and descendants thereof).
  One can note a curiousity about \lstinline$Striatum dorsal region$, which only has one descendant, 
   caudoputamen (which corresponds to the index \lstinline$uu=1$ in the above loop). The inclusion of caudoputamen 
    in the dorsal region of the striatum is therefore trivial, and 
     the two labels \lstinline$Striatum dorsal region$ and \lstinline$Caudoputamen$ can be treated as 
      synonyms of each other.\\

$\bullet$ {\bf{Example \arabic{exampleCounter}. \addtocounter{exampleCounter}{1} The \lstinline$fine$ annotation compared to the \lstinline$big12$ annotation}}.
 Let us show that the \lstinline$'fine$ annotation is a refinement of the \lstinline$big12$
 annotation. It is manifest from Table \ref{annotationSystems} that the \lstinline$fine$
 annotation covers fewer voxels than, but we can check that 1) all these voxels are also 
 in the \lstinline$big12$ annotation, and 2) each region in the \lstinline$fine$ annotation 
 intersects only one region in the \lstinline$big12$ annotation.\\
\begin{lstlisting}
% take the fine and big annotation, their ids and labels
identifierIndex = 4
cor = Ref.Coronal;
ann = cor.Annotations;
identifierIndexBig = 5;
annotBig = get_annotation( cor, ann.identifier{ identifierIndexBig  } );
idsBig = ann.ids{ identifierIndexBig };
labelsBig = ann.labels{ identifierIndexBig };  
annotFine = get_annotation( cor, ann.identifier{ identifierIndex } );
idsFine = ann.ids{ identifierIndex };
labelsFine = ann.labels{ identifierIndex };
%how are the voxels in a given fine annotation labelled in big12 
for ff = 1 : numel( idsFine )
    idFine = idsFine( ff );
    annotLoc = annotFine;
    annotLoc( annotLoc ~= idFine ) = 0;
    valsBig = unique( annotBig( find( annotFine == idFine ) ) );
    % if this region is not included in one of the big12 region,
    % valsBig contains 0 and/or several integers, and the following
    % causes an error  
    if numel( valsBig ) ~= 1 && ~isempty( find( valsBig == 0 ) ) 
        errorAtlasToBig = 'The annotation is not a refinement of big12'
        else
        indexParentBigAtlas( ff ) = find( idsBig == valsBig );
    end
end    
labelsParentBigAtlas = labelsBig( indexParentBigAtlas );
annotationFineToBig12.labelsFine = labelsFine;
annotationFineToBig12.indexParentBigAtlas = indexParentBigAtlas;
annotationFineToBig12.labelsParentBigAtlas = labelsParentBigAtlas;
\end{lstlisting}
Now from \lstinline$big12$ to \lstinline$fine$:
\begin{lstlisting}
% take the fine and big annotation, their ids and labels
identifierIndex = 4
cor = Ref.Coronal;
ann = cor.Annotations;
identifierIndexBig = 5;
annotBig = get_annotation( cor, ann.identifier{ identifierIndexBig  } );
idsBig = ann.ids{ identifierIndexBig };
labelsBig = ann.labels{ identifierIndexBig };  

annotFine = get_annotation( cor, ann.identifier{ identifierIndex } );
idsFine = ann.ids{ identifierIndex };
labelsFine = ann.labels{ identifierIndex };
for bb = 1 : numel( idsBig)
    idBig = idsBig( bb );
    valsFine = unique( annotFine( find( annotBig == idBig ) ) );
    indicesFine = [];   
    for vv = 1 : numel( valsFine )
       valFine = valsFine( vv );
       if valFine ~= 0 
           indicesFine = [ indicesFine, find( idsFine == valFine ) ];
       end
    end    
    annotationBig12ToFine.indicesInFineAtlas{ bb } = indicesFine;
    annotationBig12ToFine.subregionsInFineAtlas{ bb } = labelsFine( indicesFine );
    annotationBig12ToFine.labelBig{ bb } = labelsBig{ bb };
end
\end{lstlisting}
The toolbox contains these code snippets as 
two functions that work out the refinement
 of the \lstinline$big12$ annotation, and the organisation of the 
\lstinline$fine$ annotation into larger regions of the brain.
 One can check that the following reproduces the above results:
\begin{lstlisting}
annotationFineToBig12 = annotation_fine_to_big12( Ref );
annotationBig12ToFine = annotation_big12_to_fine( Ref );
\end{lstlisting}
For instance, the 16-th region in the \lstinline$fine$
 is Nucleus Accumbens. Check it:
\begin{lstlisting}
display( Ref.Coronal.Annotations.labels{ 4 }( 16 ) );
% check that this region is a the right index in annotationFineToBig12
display( annotationFineToBig12.labelsFine( nAccIndex ) );
% In which region of Big12 is it included?
display( annotationFineToBig12.labelsParentBigAtlas( nAccIndex ) );
% Apart from nucleus accumbens which are the other subregions of the
% striatum?
striatumIndexInBig12 = annotationFineToBig12.indexParentBigAtlas( nAccIndex );
% check that the striatum is a the right index in annotationBig12ToFine
display( annotationBig12ToFine.labelBig{ striatumIndexInBig12 } );
%the subregions of the striatum are the following
display( annotationBig12ToFine.subregionsInFineAtlas{ striatumIndexInBig12 } );
\end{lstlisting}

\chapter{Genes versus neuroanatomy}
\section{Localization scores}
\subsection{Localization scores of a single gene in the ARA}
 Let us define \cite{markerGenes}
 the localization score $\lambda_\omega(g)$ of a gene $g$ in a region
$\omega$  as the ratio of the squared
$L^2$-norm of the expression energy of  gene $g$
 in region $\omega$ to the squared
 $L^2$-norm of the expression energy of  gene $g$ in the set $\Omega$ of voxels
 that are annotated in the version of the ARA that contains $\omega$. 

\begin{equation}
\lambda_\omega(g) = \frac{\sum_{v\in\omega} E(v,g)^2 }{ \sum_{v\in\Omega} E(v,g)^2 },
\label{localizationScoreSingle}
\end{equation}
It can be computed from a voxel-by-gene matrix (with one score per column
for a fixed region $\omega$),
using the function
\lstinline$localization_from_id.m$, 
given the numerical id of the region $\omega$ and the numerical identifier 
corresponding to an annotation containing $\omega$.\\

$\bullet$ {\bf{Example \arabic{exampleCounter}. One gene, one region.\addtocounter{exampleCounter}{1}} Let us compute the localization 
 score of {\emph{Pak7}} in the cerebral cortex, as defined 
 by the \lstinline$big12$ annotation of the left hemisphere.}\\
\begin{lstlisting}
cor = Ref.Coronal;
ann = cor.Annotations;
identifierIndex = 5;
ids = ann.ids{ identifierIndex };
labels = ann.labels{ identifierIndex };
% where is the Cerebral cortex in the atlas?
cortexIndex = find( strcmp( labels, 'Cerebral cortex' ) );
idCortex = ids( cortexIndex );
indexPak7 = find( strcmp( genesAllen, 'Pak7' ) == 1 );
localizationPak7Cortex = localization_from_id( Ref,...
 E( :, indexPak7 ), identifierIndex, idCortex );
\end{lstlisting}

The function \lstinline$localization_scores_region_by_gene.m$
computes the localization scores of all the columns of the data matrix
in a given system of annotation.
 The set of voxels called
\lstinline$Basic cell groups and regions$ can be skipped by 
 choosing \lstinline$struct( 'indInit', 2 )$ as the last argument 
 of the function\\ 
\lstinline$localization_scores_region_by_gene.m$ in the
\lstinline$'big12$ annotation. The use of this function is illustrated
\addtocounter{exampleCounter}{1}.

\begin{figure}
\centering
\includegraphics[width=\figWidth,keepaspectratio]{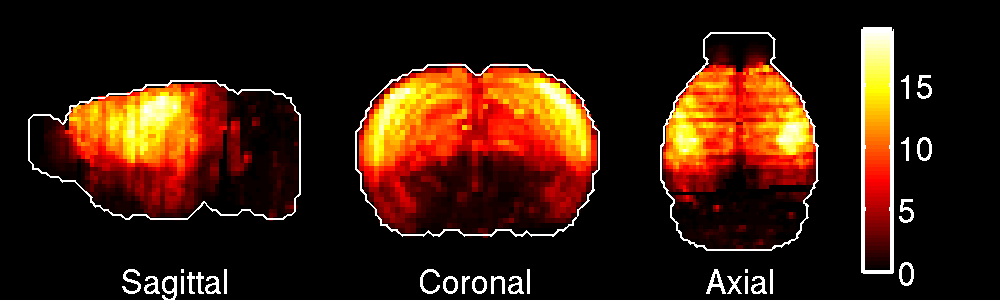}
\caption{Maximal-intensity projection of the best-localized gene in the cerebral cortex.}
\label{singleLocMarkerIdentifier5Region2}
\end{figure}

$\bullet$ {\bf{Example \arabic{exampleCounter}. More genes, more regions: 
comparison to the atlas by localization scores.\addtocounter{exampleCounter}{1}}}
The code snippet below should reproduce Figure \ref{singleLocMarkerIdentifier5Region2}
among other things.
\begin{lstlisting}
 identifierIndex = 5;
 localizationRegionByGeneBig12 = localization_scores_region_by_gene( Ref,...
  E, identifierIndex, struct( 'indInit', 1 ) );
 % check that the columns of the matrix of localization 
 % scores sum to 1
 numGenes = size( locScores, 2 );
 shouldBeOne = mean( sum( locScores ) );
 display( shouldBeOne );
 % look at the best-localized gene in each brain region
 for rr = 2 : numel( ids )
     labelReg = labels( rr );
     display( labelReg ); 
     regionScores = locScores( rr, : );
     bestGeneIndex = find( regionScores == max( regionScores ) );
     bestGeneIndex = bestGeneIndex( 1 );
     display( genesAllen( bestGeneIndex ) );
     bestExpressionByLocalization = E( :, bestGeneIndex );
     bestExpressionByLocalization =...
           make_volume_from_labels( bestExpressionByLocalization, brainFilter );
     plot_intensity_projections( bestExpressionByLocalization );
     saveas( gcf, [ 'singleLocMarkerIdentifier',...
             num2str( identifierIndex ), 'Region', num2str( rr ) ], 'png' );
     pause( 1 ); 
 end    
\end{lstlisting}

\begin{figure}
\centering
\includegraphics[width=\figWidth,keepaspectratio]{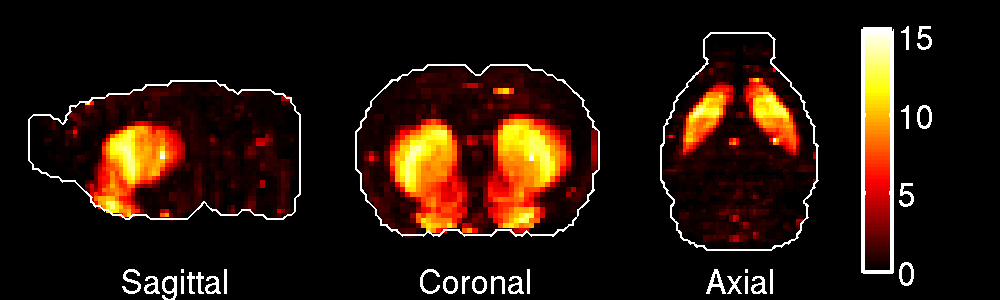}
\caption{Maximal-intensity projection of the best-localized gene in the striatum.}
\label{singleLocMarkerIdentifier5Region6}
\end{figure}


\subsection{Localization scores of sets of genes in the ARA}
Equation  \ref{localizationScoreSingle} is naturally 
generalised to a linear combination of gene-expression vectors, weighted 
 by real coefficients: 
\begin{equation}
E_\alpha( v ) : = \sum_{g = 1}^{G} \alpha_g E( v, g ),\;\;\;
 \alpha = (\alpha_1,\dots,\alpha_G) \in {\mathbf{R}}^G,
\label{eq:superposition}
\end{equation}
where $G$ is the number of genes in our dataset ($G=3,041$ by default
 when using the start-up file \lstinline$mouse_start_up.m$).\\
 Let us define the localization score in the brain region $\omega$ of a weighted set
of genes  encoded by Equation \ref{eq:superposition} as
\begin{equation}
         \lambda_\omega( \alpha ) = \frac{\sum_{v\in\omega} \left( \sum_g
  \alpha_g E( v, g )\right)^2 }{\sum_{v\in\Omega} \left( \sum_g \alpha_g
  E( v, g )\right)^2} = \frac{\alpha^t J^\omega \alpha}{\alpha^t  J^\Omega \alpha},
\label{lambdaOmega}
\end{equation}

$\bullet$ {\bf{Example \arabic{exampleCounter}. Let us compute the best
 generalized localization scores in the \lstinline$big12$ annotation.\addtocounter{exampleCounter}{1}}}\\
\begin{lstlisting}
% Consider the big12 annotation
identifierIndex = 5;
optionsGen = struct( 'recomputeQuad', 0, 'saveResults', 0, 'recomputeQuadTot', 0 );
% Cortex only
for rr = 2
% Whole atlas
% for rr = 1 : numel( Ref.Coronal.Annotations.ids{ identifierIndex } )
    localizationGeneralizedRegionBig12{ rr } = localization_generalized( Ref,...
             E, identifierIndex, rr, optionsGen );
    genEigenVector = localizationGeneralizedRegionBig12{ rr }.genVec;
    weightedSumOfGenes = E * genEigenVector;
    vol = make_volume_from_labels( weightedSumOfGenes, brainFilter );
    %plot the expression of the solution
    plot_intensity_projections( vol );
    saveas( gcf, [ 'genLocMarkerIdentifier', num2str( identifierIndex ),...
    'Region', num2str( rr ) ], 'png' );
    pause( 3 );      
end
\end{lstlisting}    
\begin{figure}
\centering
\includegraphics[width=\figWidth,keepaspectratio]{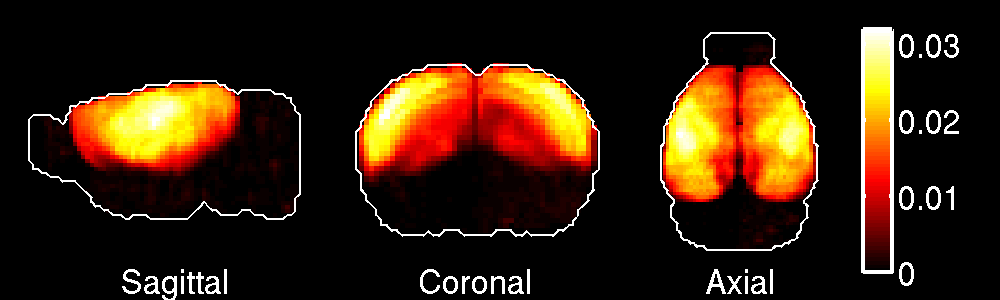}
\caption{Maximal-intensity projection of the best generalized marker if the cerebral cortex.}
\label{genLocMarkerIdentifier5Region2}
\end{figure}
\section{Fitting scores in the ARA}
\subsection{Fitting scores of a single gene in the ARA}
The fitting score $\phi_\omega(g)$ of a gene $g$ in a region
$\omega$ is defined
\begin{equation}
\phi_\omega(g) = 1 - \frac{1}{2} \sum_{v\in \Omega}\left( E^{\mathrm{norm}}_g(v ) - \chi_\omega(v)\right )^2,
\label{fittingScoreSingle}
\end{equation}
where $E^{\mathrm{norm}}_g$ is the $L^2$-normalized $g$-th column $E_g$ of the matrix of gene-expression energies:
\begin{equation}
E^{\mathrm{norm}}_g( v ) = \frac{E(v,g)}{\sqrt{\sum_{w=1}^V E( w,g)^2}},
\end{equation}
 the symbol $\Omega$ denotes the set of voxels in a given system of annotation that contains
 region $\omega$. The definition is the only decreasing affine function of the squared $L^2$
 norm of the difference between the normalized gene-expression vector of gene $g$ 
 and the characteristic function $\chi_\omega$ of the region $\omega$: 
\begin{equation}
\chi_\omega( v ) = \frac{{\mathbf{1}}(v\in\omega)}{\sqrt{\sum_{w\in\omega}{\mathbf{1}}(w\in\omega)}},
\label{characteristic}
\end{equation}
where the denominator in Equation \ref{characteristic} ensure the 
 $L^2$-normalization $\sum_{v = 1}^V \chi_\omega(v )^2 = 1$.\\

Given a gene, a system of annotation chosen among the 
ones in Table \ref{annotationSystems}, and the numerical id 
 of a region in this system of annotation, the function 
 \lstinline$fitting_from_id.m$ computes the fitting score of the gene to the corresponding region.\\

$\bullet$ {\bf{Example \arabic{exampleCounter}. One gene, 
one region in a given annotation.\addtocounter{exampleCounter}{1}}}
Let us compute the fitting score of {\emph{Pak7}} in $\omega$ = \lstinline$Cerebral cortex$,
 in the \lstinline$big12$ annotation (hence \lstinline$identifierIndex = 5$, see Table 
 \ref{annotationSystems}).\\  
\begin{lstlisting}
ann = cor.Annotations;
identifierIndex = 5;
ids = ann.ids{ identifierIndex };
labels = ann.labels{ identifierIndex };
% where is the Cerebral cortex in the atlas?
cortexIndex = find( strcmp( labels, 'Cerebral cortex' ) );
idCortex = ids( cortexIndex );
indexPak7 = find( strcmp( genesAllen, 'Pak7' ) == 1 );
fittingPak7Cortex = fitting_from_id( Ref, E( :, indexPak7 ),...
 identifierIndex, idCortex );
\end{lstlisting}

 The function \lstinline$fitting_from_id.m$ can also used to compute the 
 fitting score of several genes. If the second argument of the function 
 is a voxel-by-gene matrix with $p$ columns, the function returns an 
array of $p$ fitting scores arranged in the same order as the columns.\\

 Given a version of the ARA (specified by the index \lstinline$identifierIndex$),
 a region-by-gene matrix of all the fitting scores of all genes
 corresponding to the columns of the data matrix. Note that 
 the columns of this region-by-gene score matrix do not sum 
 to a constant (the squares of the entries of each column sum 
 to the square of the fraction of the gene-expression that projects 
 onto the set of voxels in the annotation, which is at most 1).\\

 $\bullet$ {\bf{Example \arabic{exampleCounter}. Fitting scores of
 all genes in all the regions in the \lstinline$big12$ annotation.\addtocounter{exampleCounter}{1}}}
\begin{lstlisting}
identifierIndex = 5;
ids = ann.ids{ identifierIndex };
labels = ann.labels{ identifierIndex };
fittingScoresRegionByGene = fitting_scores_region_by_gene( Ref,...
  E, identifierIndex, options );
fittingScores = fittingScoresRegionByGene.fittingScores;
for rr = 2 : numel( ids )
    labelReg = labels( rr );
    display( labelReg ); 
    fittingScoresRegionByGene = fittingScoresRegionByGeneBig12.fittingScores;
    regionScores = fittingScoresRegionByGene( rr, : );
    bestGeneIndex = find( regionScores == max( regionScores ) );
    bestGeneIndex = bestGeneIndex( 1 );
    display( genesAllen( bestGeneIndex ) );
    bestExpressionByFitting = E( :, bestGeneIndex );
    bestExpressionByFitting = make_volume_from_labels( bestExpressionByFitting,...
                        brainFilter );
    plot_intensity_projections( bestExpressionByFitting );
    pause( 1 ); 
end    
\end{lstlisting}
 
\begin{figure}
\centering
\includegraphics[width=\figWidth,keepaspectratio]{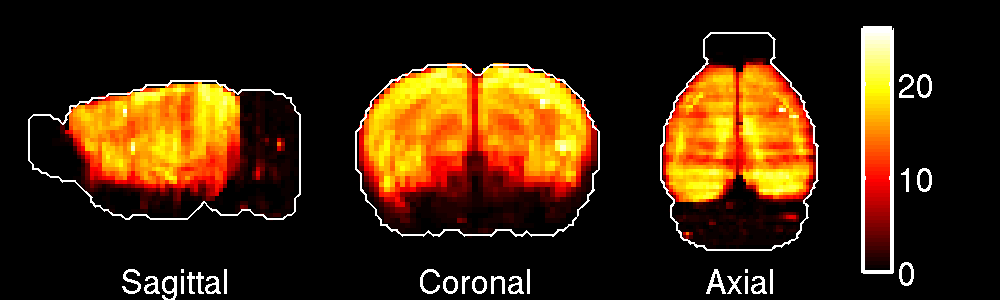}
\caption{Maximal-intensity projection of the best-fitted gene in the cerebral cortex.}
\label{singleFitMarkerIdentifier5Region2}
\end{figure}

\begin{figure}
\centering
\includegraphics[width=\figWidth,keepaspectratio]{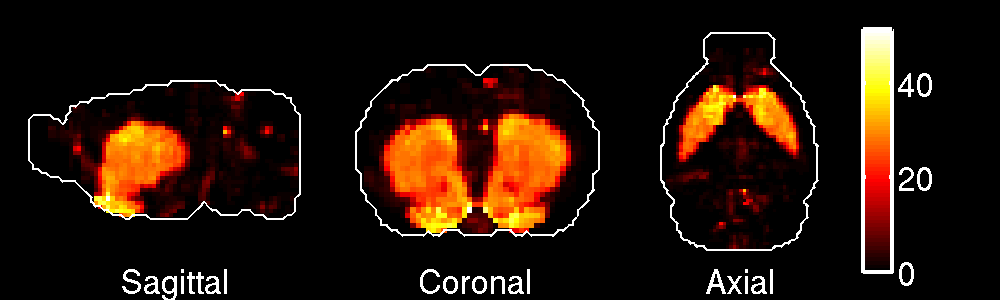}
\caption{Maximal-intensity projection of the best-fitted gene in the striatum.}
\label{singleFitMarkerIdentifier5Region6}
\end{figure}

\subsection{Fitting scores of sets of genes in the ARA}
Like the localization score, the fitting score can be generalized to 
linear combinations of sets of genes:
\begin{equation}
\phi_\omega(\alpha)  = 1 - \frac{1}{2} \sum_{v\in \Omega}\left( E^{\mathrm{norm}}_\alpha( v )
 - \chi_\omega( v ) \right)^2,\;\;\; \alpha= (\alpha_1,\dots,\alpha_G) \in {\mathbf{R}}_+^G,
\label{genFitting}  
\end{equation}
where $E^{\mathrm{norm}}_\alpha$ is the $L^2$-normalized gene-expression vector 
corresponding to the coefficients $(\alpha_1,\dots,\alpha_G)$
\begin{equation}
 E^{\mathrm{norm}}_\alpha( v ) = \frac{E_\alpha( v )}{\sqrt{\sum_{v\in\Omega}E_\alpha(v)^2}}.
\end{equation}

\begin{equation}
{\mathrm{ErrFit}}^{\omega,\Lambda}_{L^1-L^2}( \{\alpha\}) = || E^{\mathrm{norm}}_\alpha - \chi_\omega||^2_{L^2} + \Lambda||\alpha||_{L^1},
\label{eq:L1L2}
\end{equation}
which can be minimized wrt the weights of the genes using Matlab code implementing 
 an interior-point method by Koh \cite{L1L2}:
\begin{equation}
\alpha_\omega^\Lambda= {\mathrm{argmin}}_{\alpha \in {\mathbf{R}}_+^G}{\mathrm{ErrFit}}^{\omega,\Lambda}_{L^1-L^2}( \{\alpha\}).
\label{eq:optimPos}
\end{equation}

\begin{figure}
\centering
\includegraphics[width=\figWidth,keepaspectratio]{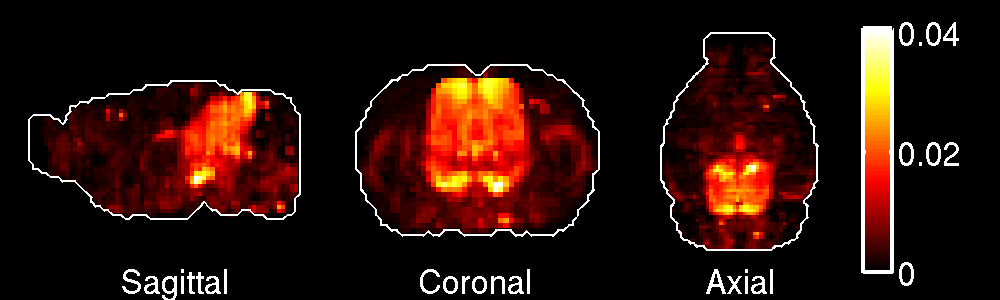}
\caption{Maximal-intensity projection of the best-fitted set of genes (with weights of both signs) in the midbrain (constructed 
 out of the best-fitted 100 genes)}
\label{setSignedFitMarkerIdentifier5Region10}
\end{figure}

\begin{figure}
\centering
\includegraphics[width=\figWidth,keepaspectratio]{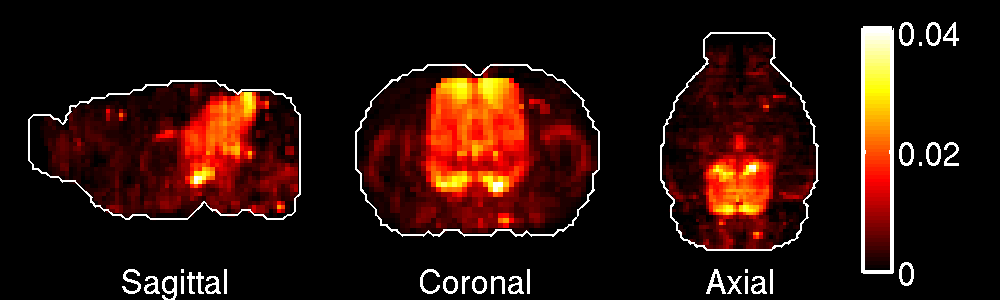}
\caption{Maximal-intensity projection of the best-fitted set of genes (with positive weights) in the midbrain (constructed 
 out of the best-fitted 100 genes).}
\label{setPositiveFitMarkerIdentifier5Region10}
\end{figure}

$\bullet$ {\bf{Example \arabic{exampleCounter}. Best-fitted sets of 
 genes in (some regions of) the \lstinline$big12$ annotation.\addtocounter{exampleCounter}{1}}}
The code below should reproduce Figures \ref{setSignedFitMarkerIdentifier5Region10} and \ref{setPositiveFitMarkerIdentifier5Region10}
\begin{lstlisting}
optionsSet = struct( 'lambdaVals', [ 0.005 ],...
      'identifierIndex', 5, 'numGenes', 20 );
lambdaVals = optionsSet.lambdaVals;
numGenes = optionsSet.numGenes;
numLambdaVals = numel( lambdaVals );
cor = Ref.Coronal;
ann = cor.Annotations;
identifierIndex = optionsSet.identifierIndex;
ids = ann.ids{ identifierIndex };
labels = ann.labels{ identifierIndex };
annot = get_annotation( cor, ann.identifier{ identifierIndex } );
%focus on midbrain
anatInds = 10;
display( numGenes )
optionsSigned = struct( 'identifierIndex', identifierIndex, 'numGenesKept', numGenes,...
        'regionIndexInit', 2, 'lambdaMult', 0, 'saveResults', 0,...
        'verboseExec', 1, 'indsInAtlas', anatInds, 'positiveConstraint', 0 );
optionsPos = struct( 'identifierIndex', identifierIndex, 'numGenesKept', numGenes,...
        'regionIndexInit', 2, 'lambdaMult', 0, 'saveResults', 0,...
        'verboseExec', 1, 'indsInAtlas', anatInds, 'positiveConstraint', 1 );
for vv = 1 : numLambdaVals
    lambdaCurrent = lambdaVals( vv )
    if lambdaCurrent > 0
       optionsSigned.lambdaMult = lambdaCurrent;
       optionsPos.lambdaMult = lambdaCurrent;
       signedFitting = signed_fitting( Ref, E, optionsSigned );
       positiveFitting = signed_fitting( Ref, E, optionsPos );
       exploreL1L2ParameterSpaceLoc = struct( 'signedFitting', signedFitting,...
                  'positiveFitting', positiveFitting );
       exploreL1L2ParameterSpace{ vv } = exploreL1L2ParameterSpaceLoc;
       for aa = 1 : numel( anatInds )
           anatInd = anatInds( aa ); 
           signedFittingCoeffs = signedFitting.coeffsReg{ anatInd };
           numGenesKept = signedFitting.numGenesKept;
           signedFittingIndsKept = signedFitting.fittingInds{ anatInd };
           signedFittingIndsKept = signedFittingIndsKept( 1 : numGenesKept );
           signedFittingSol = E( :, signedFittingIndsKept ) *...
                                  signedFittingCoeffs;
           signedFittingSol = make_volume_from_labels( signedFittingSol,...
                         brainFilter );
           plot_intensity_projections( signedFittingSol );
           pause( 2 );
           positiveFittingCoeffs = positiveFitting.coeffsReg{ anatInd };
           numGenesKept = signedFitting.numGenesKept;
           positiveFittingIndsKept =...
                   positiveFitting.fittingInds{ anatInd }( 1 : numGenesKept );
           positiveFittingSol = E( :, positiveFittingIndsKept ) *...
                                    signedFittingCoeffs;
           positiveFittingSol = make_volume_from_labels( positiveFittingSol,...
                    brainFilter );
           plot_intensity_projections( positiveFittingSol );
           pause( 2 );
         end
    end
end
\end{lstlisting}

\chapter{Genes versus genes: co-expression networks}
\section{Co-expression networks of genes in the Allen Atlas}
The co-expression of two genes in the Allen Atlas
is defined as the cosine similarity between their gene-expression 
vectors in voxel space. Given a voxel-by-gene matrix
 containing the brain-wide expression energies (as in Equation \ref{ExpressionEnergy}),
 the corresponding gene-by-gene matrix of co-expressions of the full set
of genes, or ${\mathrm{coExpr}}^{\mathrm{full}}$, is the symmetric matrix 
 with entries equal to the co-expression of pairs of genes, as in Equation 
\ref{coExpressionFull}.\\

\begin{equation}
{\mathrm{coExpr}}^{\mathrm{full}}( g, g'):= \frac{\sum_{v = 1}^V  E( v, g ) E( v, g' )}
{\sqrt{\sum_{u= 1}^VE(u,g )^2\sum_{w= 1}^VE(w,g' )^2}}.
\label{coExpressionFull}
\end{equation}

The matrix ${\mathrm{coExpr}}^{\mathrm{full}}$ can be computed as follows 
in Matlab:\\
\begin{lstlisting}
cor = Ref.Coronal;
% divide each column of the data matrix by its L2-norm
ENormalised = normalise_integral_L2( E );
% co-expression matrix of the full set of genes 
coExpressionFull = ENormalised' * ENormalised;
\end{lstlisting}
The function  \lstinline$co_expression_matrix.m$ takes a voxel-by-gene 
 matrix as an argument and returns the gene-by-gene co-expression 
matrix defined by Equation \ref{coExpressionFull}, which equals 
 the matrix \lstinline$coExpressionFull$ defined in the above code snippet if the 
 full voxel-by-gene matrix of gene expression  energies is used as an argument.
 Other versions of the Allen Atlas than the brain-wide standard annotations
 can be specified in the options to restrict the voxels to one of the 
 annotations described in Table \ref{annotationSystems}.

\begin{figure}
\centering
\includegraphics[width=\figWidth,keepaspectratio]{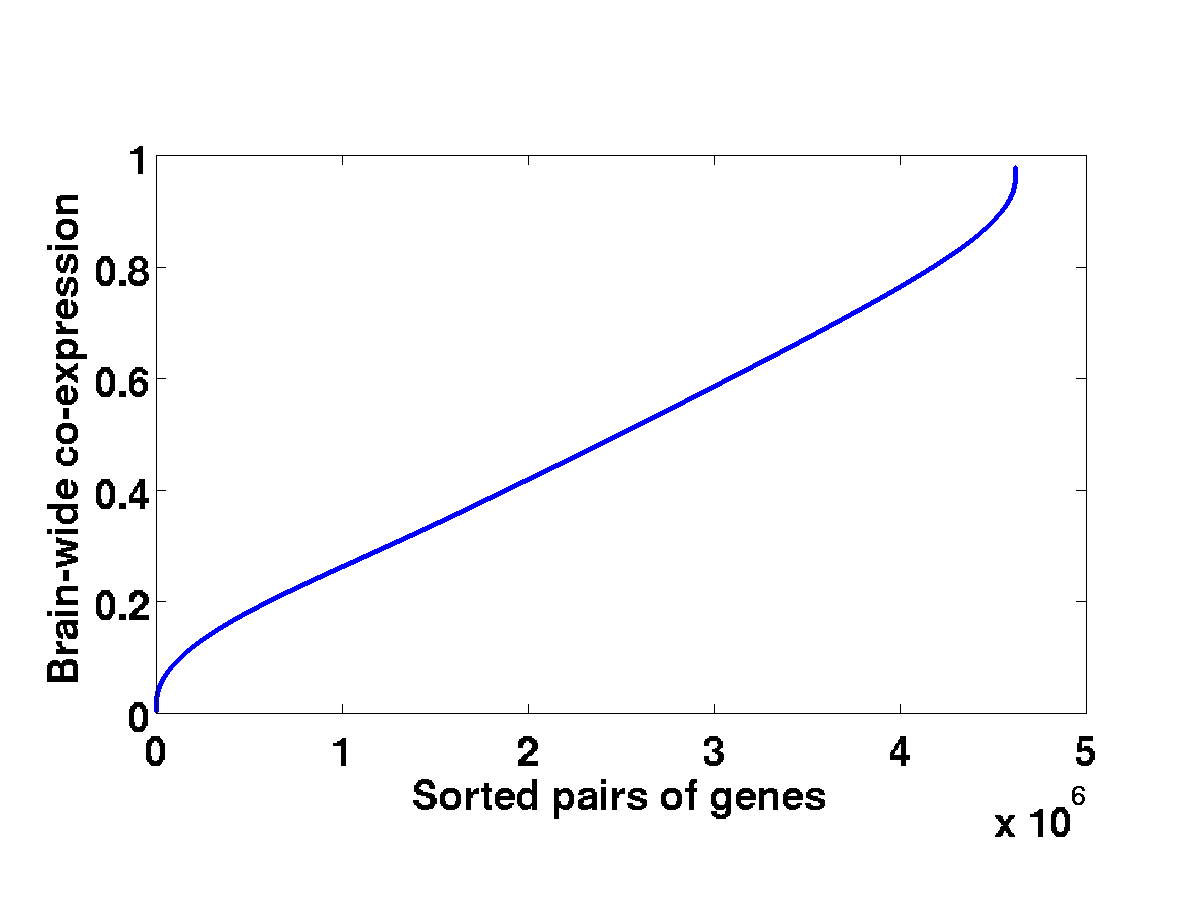}
\caption{Sorted upper-diagonal elements of the brain-wide co-expression matrix.}
\label{sortedBrainWideCoExpressions}
\end{figure}

\begin{figure}
\centering
\includegraphics[width=\figWidth,keepaspectratio]{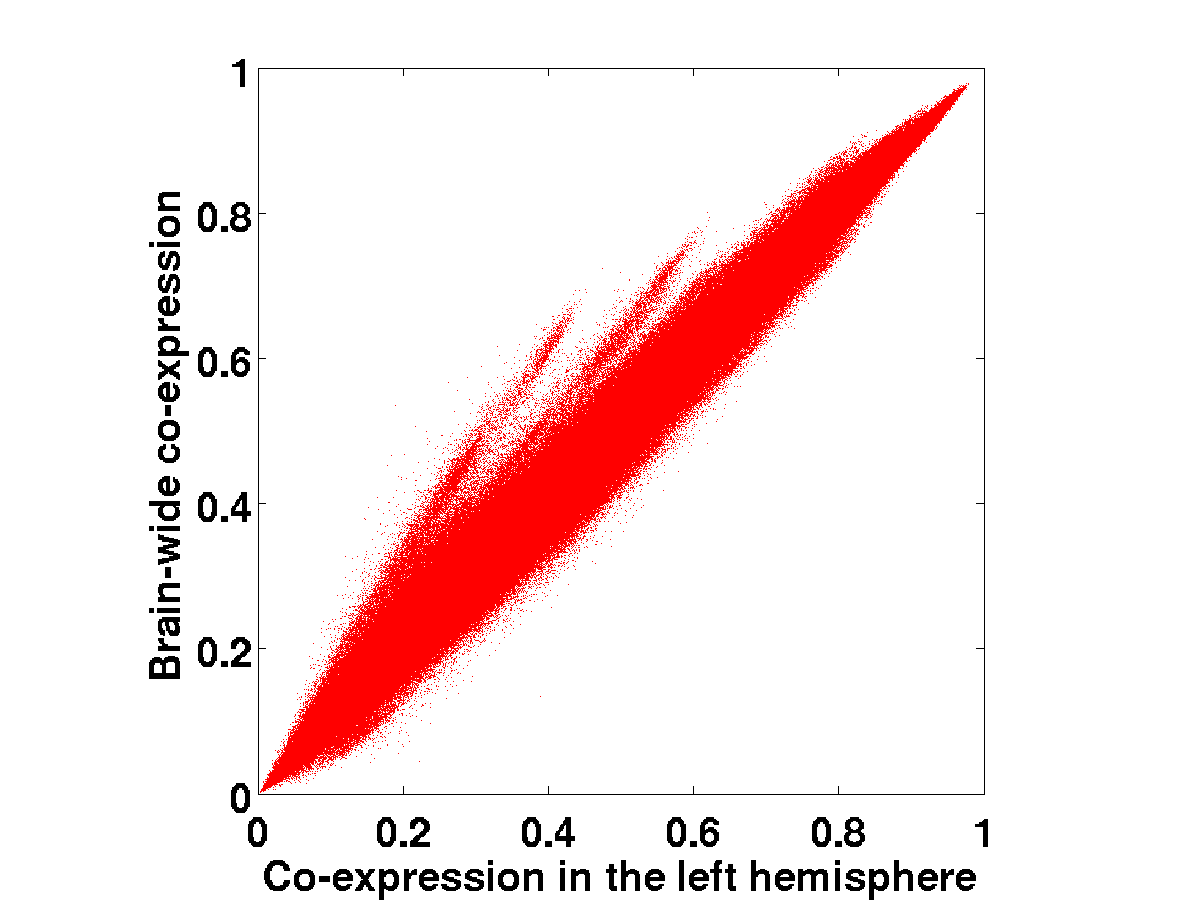}
\caption{Upper-diagonal elements of the  co-expression matrix in the left hemisphere
 plotted  against upper-diagonal elements of the brainwide co-expression matrix. 
The deviation from the diagonal reflects the difference between the sets of 
voxels in the \lstinline$big12$ and \lstinline$standard$ annotations}
\label{coExpressionLeftVersusWhole}
\end{figure}

$\bullet$ {\bf{Example \arabic{exampleCounter}. Distribution of co-expression coefficients.\addtocounter{exampleCounter}{1}}} 
The diagonal elements of the co-expression 
matrix equal 1 by construction, and the co-expression matrix is symmetric. Hence, 
the distibution of co-expression coefficients in the atlas is given by the upper diagonal coeefficients of the co-expression matrix, which can be extracted using the function 
  \lstinline$upper_diagonal_coeffs.m$, as in the code snipped below, which plots the 
distribution of brain-wide co-expression coefficients (Figure \ref{sortedBrainWideCoExpressions})) and compares it to the one of 
co-expression coefficients in the left hemisphere (as defined in the \lstinline$big12$ 
annotation, Figure \ref{coExpressionLeftVersusWhole}).\\
\begin{lstlisting}
coExpressionFull = co_expression_matrix( Ref, E );
% restrict the gene-expression data to the voxels that are 
% in the big12 annotation 
 optionsLeft = struct( 'identifierIndex', 5 );
% cosine distances between the gene-expression vectors of genes
% in the left hemisphere (voxels annotated in the big12 version of the ARA)
coExpressionLeft= co_expression_matrix( Ref, E, optionsLeft ); 
% consider only the non-trivial co-expressions
upperCoeffsFull = upper_diagonal_coeffs( coExpressionFull );
upperCoeffsLeft = upper_diagonal_coeffs( coExpressionLeft );
% sort the brain-wide co-expression coefficients 
[ valsFull, inds ] = sort( upperCoeffsFull );
% plot the brain-wide co-expression coefficients
coFigure = figure( 'Color', 'w', 'InvertHardCopy', 'off',...
 'Position', [ 200, 200, 800, 600 ] );
plot( valsFull, '.b', 'markersize', 4 );
xlabel( 'Sorted pairs of genes', 'fontsize', 20, 'fontweight', 'b' );
ylabel( 'Brain-wide co-expression', 'fontsize', 20, 'fontweight', 'b' ); 
set( gca, 'YLim', [ 0, 1 ], 'dataAspectRatio', [ 3*10^6 1  1 ],...
 'fontsize', 20, 'fontweight', 'b' );
 pause;
hold off; 
close all;
% plot upperCoeffsLeft against upperCoeffsFull
figure( 'Color', 'w', 'InvertHardCopy', 'off',...
 'Position', [ 200, 200, 800, 800 ] );
plot( valsFull, upperCoeffsLeft( inds ), '.r', 'markersize', 1 );
xlabel( 'Co-expression in the left hemisphere',... 
'fontsize', 20, 'fontweight', 'b' );
ylabel( 'Brain-wide co-expression', 'fontsize', 20,...
 'fontweight', 'b' ); 
set( gca, 'YLim', [ 0, 1 ], 'dataAspectRatio', [ 1 1 1 ],...
 'fontsize', 20, 'fontweight', 'b' );
\end{lstlisting}

\section{Monte Carlo analysis of brain-wide co-expression networks}

\subsection{Special sets of genes versus full atlas}
The gene-by-gene matrix ${\mathrm{coExpr}}^{\mathrm{full}}$ defines
a universe in which we would like to study 
co-expression networks of special sets of genes, in a probabilistic way.
\begin{equation}
{\mathrm{coExpr}}^{\mathrm{full}} \in {\mathcal{M}}_{G_{\mathrm{full}}}({\mathbf{R}}), 
\end{equation}
Given a set of $G_\mathrm{special}$ genes of interest, corresponding to the
 column indices $(g_1,\dots,g_\mathrm{special})$ in the data matrix, their
 co-expression matrix $\mathrm{coExpr}^{\mathrm{special}}$ is obtained by extracting the 
 submatrix of ${\mathrm{coExpr}}^{\mathrm{full}}$ corresponding to these
  indices (see Equation \ref{coExpressionSpecial}).
\begin{equation}
{\mathrm{coExpr}}^{\mathrm{special}} \in {\mathcal{M}}_{G_{\mathrm{special}}}({\mathbf{R}}), 
\end{equation}
\begin{equation}
{\mathrm{coExpr}^{\mathrm{special}}}(i,j)= {\mathrm{coExpr}}^{\mathrm{full}}(g_i, g_j),\;\; i,j 
\in [1..G_\mathrm{special}].
\label{coExpressionSpecial}
\end{equation}

\subsection{Cumulative distribution function of co-expression coefficients
 in sets of genes drawn from the Allen Brain Atlas}
 Having observed (Figure \ref{sortedBrainWideCoExpressions}) that the distribution 
 of pairwise co-expression coefficients of genes in the whole coronal 
 atlas is roughly linear in a large domain of co-expression,
  we can study the cumulative distribution
function of the co-expression coefficients in the special set, and compare it 
to the one resulting from random sets of genes (with the same number of genes
 as the special set, in order to eliminate the sample-size bias).\\

 These cumulative distribution functions are evaluated in the following way.
Again let $G_\mathrm{special}$ denote the size of the matrix ${\mathrm{coExpr}^{\mathrm{special}}}$,
 i.e. the number of genes from which ${\mathrm{coExpr}^{\mathrm{special}}}$ was computed. Consider
the set of entries above the diagonal of ${\mathrm{coExpr}^{\mathrm{special}}}$
 above the diagonal (which are the meaningful quantities in ${\mathrm{coExpr}^{\mathrm{special}}}$):
\begin{equation}
C^{\mathrm{special}} = \left\{ \mathrm{coExpr}^{\mathrm{special}}(g,h),\; 1\leq g \leq G_\mathrm{special},\;h > g \right\}.
\label{coExpressionSpecialSet}
\end{equation}
 The elements of this set are numbers between 0 and 1. 
For every number between
0 and 1, the cumulative distribution function (c.d.f.) of  $C^{\mathrm{special}}$, 
denoted by ${\mathrm{cdf}}^C$ is defined as the fraction of the elements of 
$C^{\mathrm{special}}$ that are smaller than this number:

\begin{align}
 \nonumber &{\mathrm{cdf}}^{\mathrm{special}}: [ 0,1 ] \rightarrow [ 0,1 ]\\
 &x \mapsto \frac{1}{\left|C^{\mathrm{special}}\right|} \sum_{c \in C^{\mathrm{special}}} {\mathbf{1}}(c\leq x),
\end{align}

where $\left|C^{\mathrm{special}}\right| = G_\mathrm{special}(G_\mathrm{special}-1)/2$.\\

For any set of genes, ${\mathrm{cdf}}^{\mathrm{special}}$ is a growing 
 function ${\mathrm{cdf}}^{\mathrm{special}}(0) = 0$ and
  ${\mathrm{cdf}}^{\mathrm{special}}(1) = 1$.
  For highly co-expressed genes, the growth of ${\mathrm{cdf}}^{\mathrm{special}}$ 
  is concentrated at high values of the argument (in the limit where all the genes 
 in the special set have the same brain-wide expression vector, all the entries 
 of the co-expression matrix go to $1$ and the cumulative
 distribution function converges to a Dirac measure supported at $1$). To 
  compare the function ${\mathrm{cdf}}^{\mathrm{special}}$ to what could be 
 expected by chance, let us draw $R$ random sets of $G_{\mathrm{special}}$ genes 
 from the Atlas, compute their co-expression network
 by extracting the corresponding entries from the full co-expression 
 matrix of the atlas (${\mathrm{coExpr}}^{\mathrm{full}}$). This induces
 a family of $R$ growing functions ${\mathrm{cdf}}_{i}, 1\leq\i\leq R$
 on the interval $[0,1]$
\begin{align}
 \nonumber{\mathrm{cdf}}_{i}&: [0,1] \rightarrow [0,1], 1\leq i \leq R\\
 {\mathrm{cdf}}_{i}(0)& = 0,\;\;\;{\mathrm{cdf}}_{i}(1) = 1.
\end{align}
From this family of functions, we can estimate a mean cumulative
 distribution function $\langle{\mathrm{cdf}}\rangle$ of the co-expression of sets of 
 $G_{\mathrm{special}}$ genes drawn from the Allen Atlas, by taking 
 the mean of the values of ${\mathrm{cdf}}_{i}$ across the 
  random draws:\\
\begin{equation}
\forall x \in [ 0,1 ],\;\;\; \langle{\mathrm{cdf}}\rangle(x) = \frac{1}{R}\sum_{i=1}^R{\mathrm{cdf}}_{i}(x).
\end{equation}
Standard deviations $\mathrm{cdf}^{\mathrm{dev}}$ of the distribution of c.d.f.s are estimated in the same
way on the interval $[0,1]$ (which the user of the code can discretize 
 into a regular grid using the \lstinline$coExprStep$ component of the options
of the function \lstinline$cumul_co_expr.m$, see example below):\\
 
\begin{equation}
\forall x \in [ 0,1], \;\;\;\mathrm{cdf}^{\mathrm{dev}}(x) = 
\sqrt{\frac{1}{R}\sum_{i=1}^R\left({\mathrm{cdf}}_{i}(x)-\langle{\mathrm{cdf}}\rangle(x)\right)^2}.
\end{equation}

The functions $\langle{\mathrm{cdf}}^{\mathrm{special}}\rangle$,
 $\langle{\mathrm{cdf}}\rangle$ and $\mathrm{cdf}^{\mathrm{dev}}$
 are fields of the output of the function \lstinline$cumul_co_expr.m$
whose usage is illustrated in the example below.\\

\begin{figure}
\includegraphics[width=\figWidth,keepaspectratio]{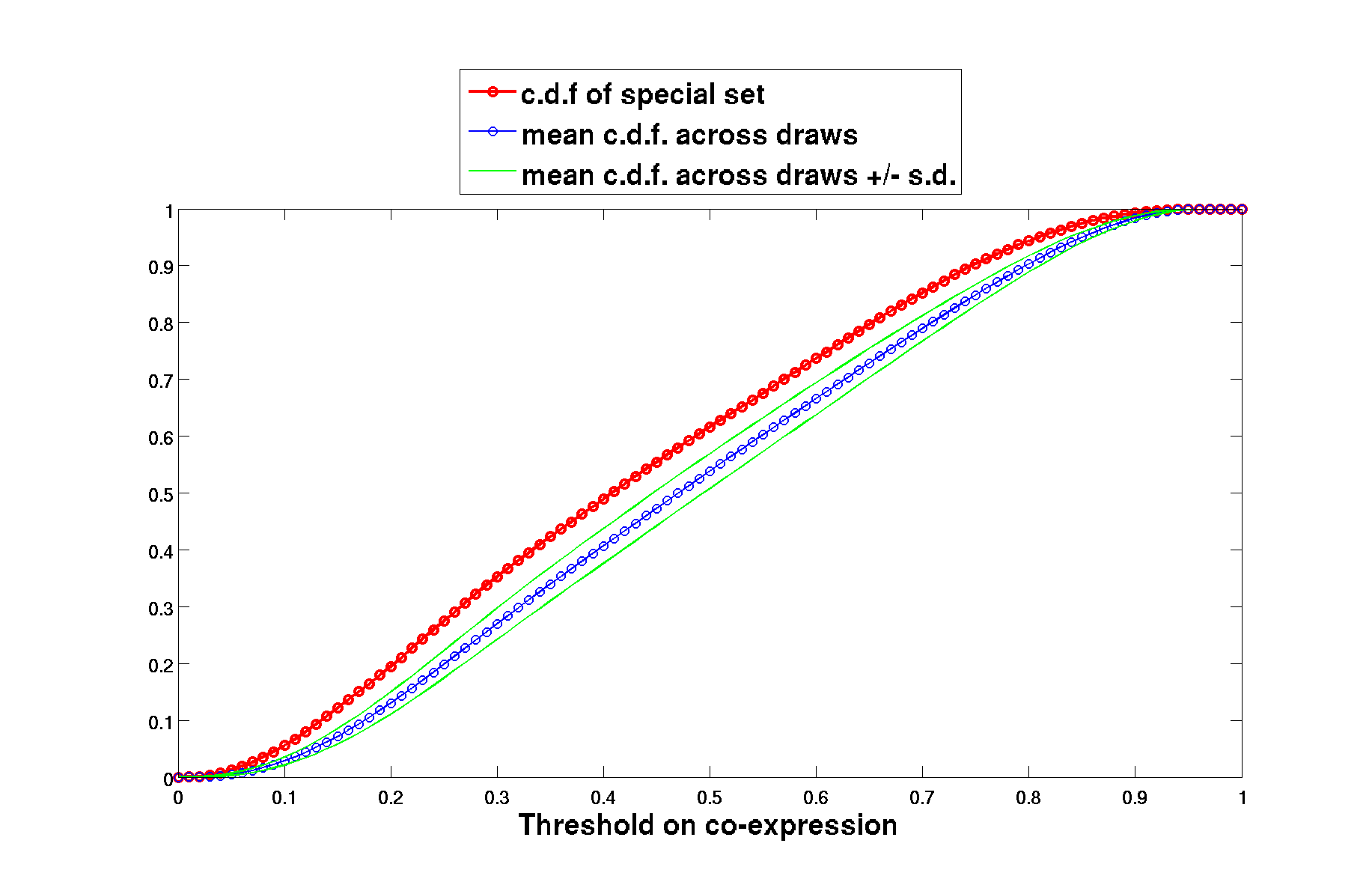}
\caption{Cumulated distribution functions of the upper-diagonal entries of the co-expression matrix
 of the special set of genes listed in \lstinline$nicotineGenes.mat$.}
\label{cumulCoExprFunction}
\end{figure}

{\bf{Example \arabic{exampleCounter}. Cumulative distribution function of
 co-expression of a special set of genes.\addtocounter{exampleCounter}{1}}}
Consider the set of 288 genes from the NicSNP database, whose position in the data matrix is encoded
 in the Matlab file \lstinline$nicotineGenes.mat$. NB: the code snippet below 
 can be applied to any special set of genes upon changing the variable \lstinline$indsSmall$.
It should reproduce Figure \ref{cumulCoExprFunction}.
\begin{lstlisting}
load( 'nicotineGenes.mat' );
indsSmall = nicotineGenes.nicotineIndicesInTop75;
% extract the special co-expression network
coExpressionSmall = coExpressionFull( indsSmall, indsSmall );
% compute the c.d.f, and simulate the distribution of 
% c.d.f.s across numPaths random sets of genes of size numel( indsSmall )
numPaths = 1000;
optionsCDF = struct( 'numPaths', numPaths, 'coExprStep', 0.01,...
       'minCoExpr', 0, 'maxCoExpr', 1, 'wantPlots', 1, 'species', 'mouse' ); 
cumulCoExpr = cumul_co_expr( coExpressionFull, coExpressionSmall, optionsCDF );
% plot the results
optionsCDF = struct( 'savePlots', 1, 'fontSize', 18,...
 'figurePosition', [200 200 1100 700], 'markerSize', 14,...
              'lineWidth', 2, 'fileName', 'cumulCoExprFunction' ); 
cumulCoExprPlot = cumul_co_expr_plot( cumulCoExpr, optionsCDF );
\end{lstlisting}

\subsection{Thresholding the  co-expression matrix}
The co-expression matrix ${\mathrm{coExpr}^{\mathrm{special}}}$ 
 corresponding to a special subset of the genes in the Allen Atlas (Equation \ref{coExpressionSpecial})
 is symmetric, like 
${\mathrm{coExpr}}^{\mathrm{full}}$,
and its entries are in the interval $[0,1]$.
 It can be mapped to a weighted graph in the following way 
(see \cite{coExpressionStats} for details). 
 The vertices
of the graph are the genes, and the edges are as follows:\\
- genes $g$ and $g'$ are linked by an edge if their co-expression is strictly positive.\\
- If an edge exists, it has weight ${\mathrm{coExpr}^{\mathrm{special}}}_{gg'}$.\\
 
 Let us define the following thresholding procedure on co-expression
graphs: given a threshold $\rho$ between $0$
and $1$,  put to zero all the
 entries of ${\mathrm{coExpr}^{\mathrm{special}}}$ that are lower than this
coefficient. The underlying graph is obtained by taking the graph 
corresponding to ${\mathrm{coExpr}^{\mathrm{special}}}$, and cutting all the links
with weight below $\rho$.
\begin{equation}
{\mathrm{coExpr}^{\mathrm{special}}}_\rho(g,h) = {\mathrm{coExpr}^{\mathrm{special}}}(g,h) 
\times\mathbf{1}\left( {\mathrm{coExpr}^{\mathrm{special}}}(g,h)  \geq \rho \right).
\label{threshEq}
\end{equation}

The more-co-expressed a set of genes is, the 
 larger the connected components of the thresholded graphs 
uderlying ${\mathrm{coExpr}^{\mathrm{special}}}_\rho$ will be,
 for any value of the threshold $\rho$.
 For instance we can study the average size of connected components
of thresholded co-expression matrices
and the size of the largest connected component 
as a function of the threshold $\rho$ :
\begin{equation}
\mathcal{A}( \rho ) = \frac{\sum_{k = 1}^G k  N_\rho(k)}{\sum_{k = 1}^G  N_\rho(k) },
\label{averageSize}
\end{equation}
\begin{equation}
\mathcal{M}( \rho ) = \mathrm{max}\left\{k \in [1..G],  N_\rho(k) > 0 \right\},
\label{maximumSize}
\end{equation}
where $N_\rho(k)$ is the number of connected components with size
$k$. The connected components are worked out using the 
 implementation of Tarjan's algorithm \cite{Tarjan} in the Matlab function
 \lstinline$graphconncomp.m$.\\

\subsection{Statistics of sizes of connected components}

 For any quantity worked out from the special co-expression 
 matrix co-expression matrix ${\mathrm{coExpr}^{\mathrm{special}}}$
 defined in Equation \label{coExpressionSpecial}, we can simulate 
 its probability distribution by repeatedly drawing random 
 random sets of genes from the atlas, 
  and recomputing the same quantity on for this set.\\

 Let us use the above-defined thresholding procedure to study a set of
 $G_{\mathrm{special}}=288$ genes obtained by intersecting the NicSNP
 database \cite{NicSNP} with the set of $G=3,041$ genes given
 by\\ \lstinline$get_genes( Ref.Coronal, 'top75CorrNoDup', 'allen' )$.
 We would like to acertain whether this set of genes is more
 co-expressed than expected by chance for a set of this size taken
 from \lstinline$genesAllen$. At each value of a regular 
 grid the threshold $\rho$ between zero and 1, the function
  \lstinline$co_expression_island_bootstrap.m$ computes 
 the maximal size and average size of connected components
 of the thresholded co-expression graph, and draws $R$ random
 sets of genes of size $G_{\mathrm{special}}$ from the atlas.
  This induces a distribution of $R$ partitions of sets 
of $G_{\mathrm{special}}$ genes into connected components,
 obtained by applying Tarjan's algorithm to each of the 
 $R$ sets of genes.\\

{\bf{Example \arabic{exampleCounter}.\addtocounter{exampleCounter}{1}}} 
Consider the set of 288 genes from the NicSNP database, whose position in the data matrix is encoded
 in the Matlab file \lstinline$nicotineGenes.mat$.
\begin{lstlisting}
% compute the full co-expression matrix coExpressionFull
coExpressionFull = co_expression_matrix( E );
load( 'nicotineGenes.mat' );
indsSmall = nicotineGenes.nicotineIndicesInTop75;
coExpressionSmall = coExpressionFull( indsSmall, indsSmall );
% Monte Carlo analysis of the graph underlying the co-expression matrix
numDraws = 1000;
optionsBootstrap = struct( 'thresholdInit', 1, 'thresholdStep', thresholdStep,...
 'numDraws', numDraws, 'thresholdFinal', min( min( coExpressionFull ) ) );
coExpressionComponentsBootStrap =...
    co_expression_components_bootstrap( coExpressionFull,...
     coExpressionSmall, optionsBootstrap );
% reproduce the three plots of this example and save them                       
optionsPlot =  struct( 'savePlots', 1,...
'fontSize', 16, 'figurePosition', [200 200 1100 700], 'markerSize', 14 );                 
fileNames = { 'coExpressionSet', 'coExpressionProba', 'coExpressionSDs' };               
coExpressionComponentsBootStrapPlot =...
 co_expression_components_bootstrap_plot(
coExpressionComponentsBootStrap, optionsPlot, fileNames );
\end{lstlisting}

\begin{figure}
\centering
\includegraphics[width=\figWidth,keepaspectratio]{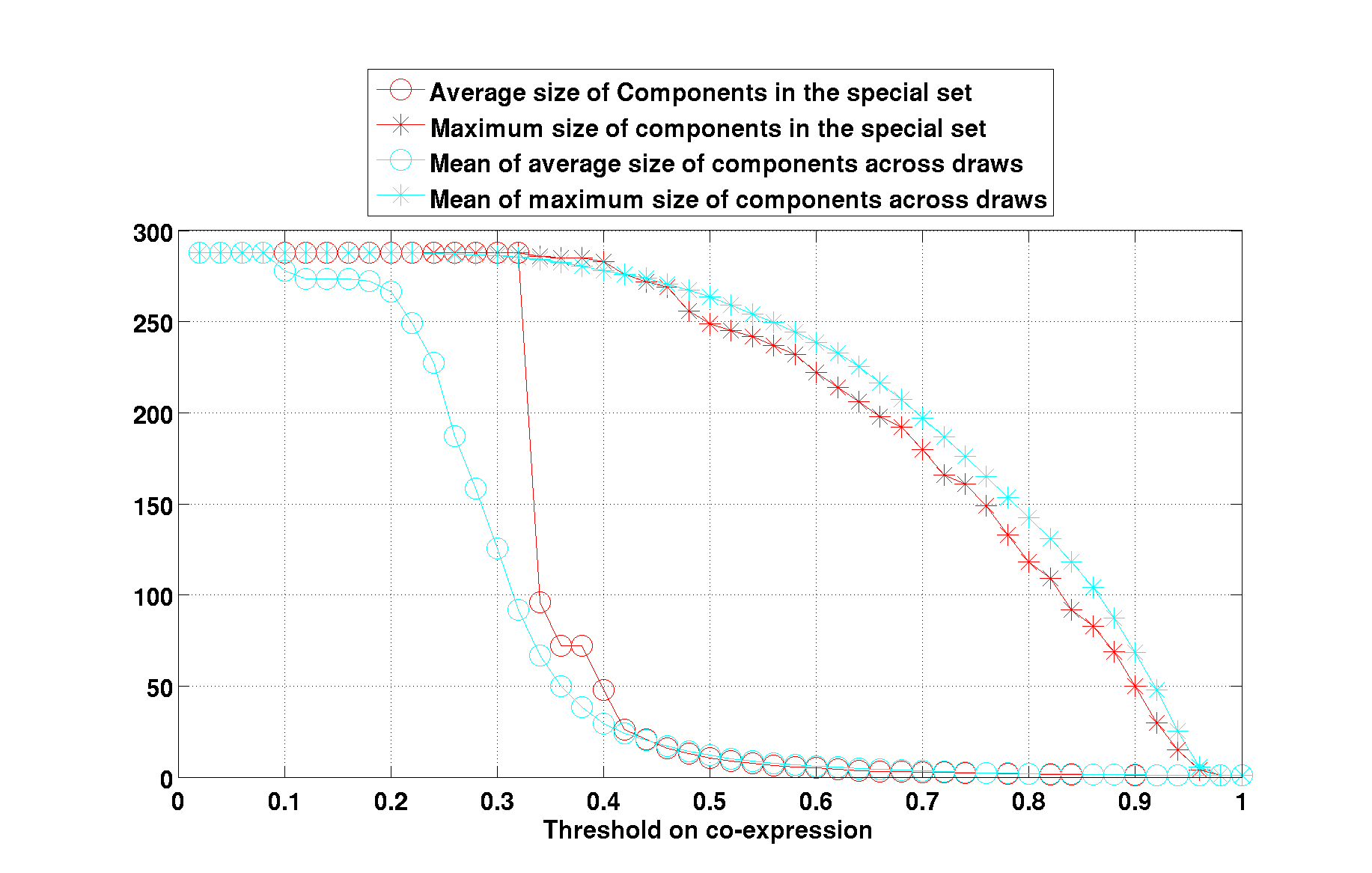}
\caption{{\bf{Monte Carlo analysis of the graph underlying the co-expression matrix of 288 genes 
 from the NicSNP database.}} Average and maximum size of connected components as a function of the threshold.}
\label{coExpressionSet}
\end{figure} 

\begin{figure}
\centering
\includegraphics[width=\figWidth,keepaspectratio]{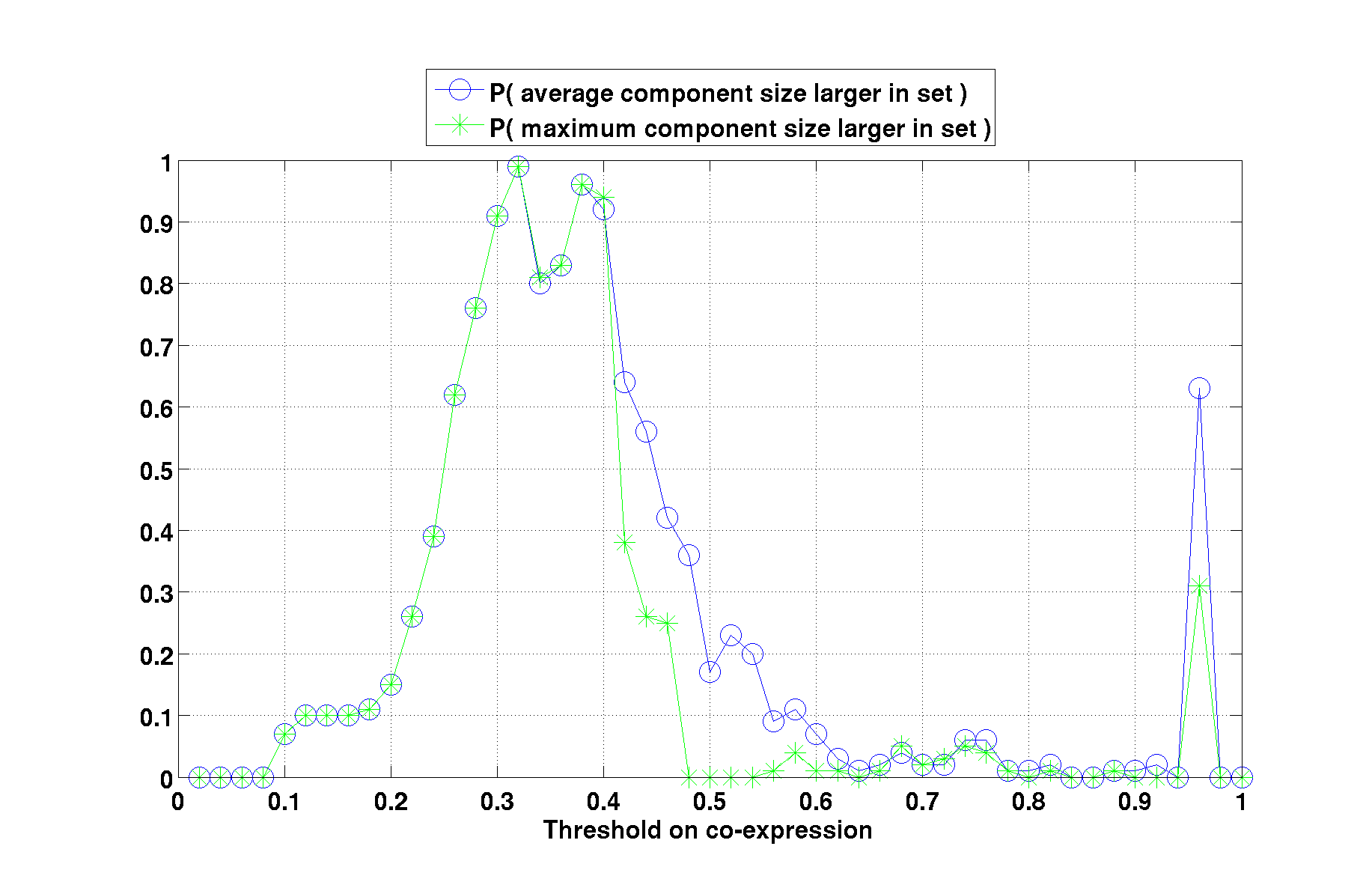}
\caption{{\bf{Monte Carlo analysis of the graph underlying the co-expression matrix of 288 genes 
 from the NicSNP database.}} Estimated probabilities for the average and maximum size of connected components to be larger than in random sets of genes of the same size.}
\label{coExpressionProba}
\end{figure}

\begin{figure}
\centering
\includegraphics[width=\figWidth,keepaspectratio]{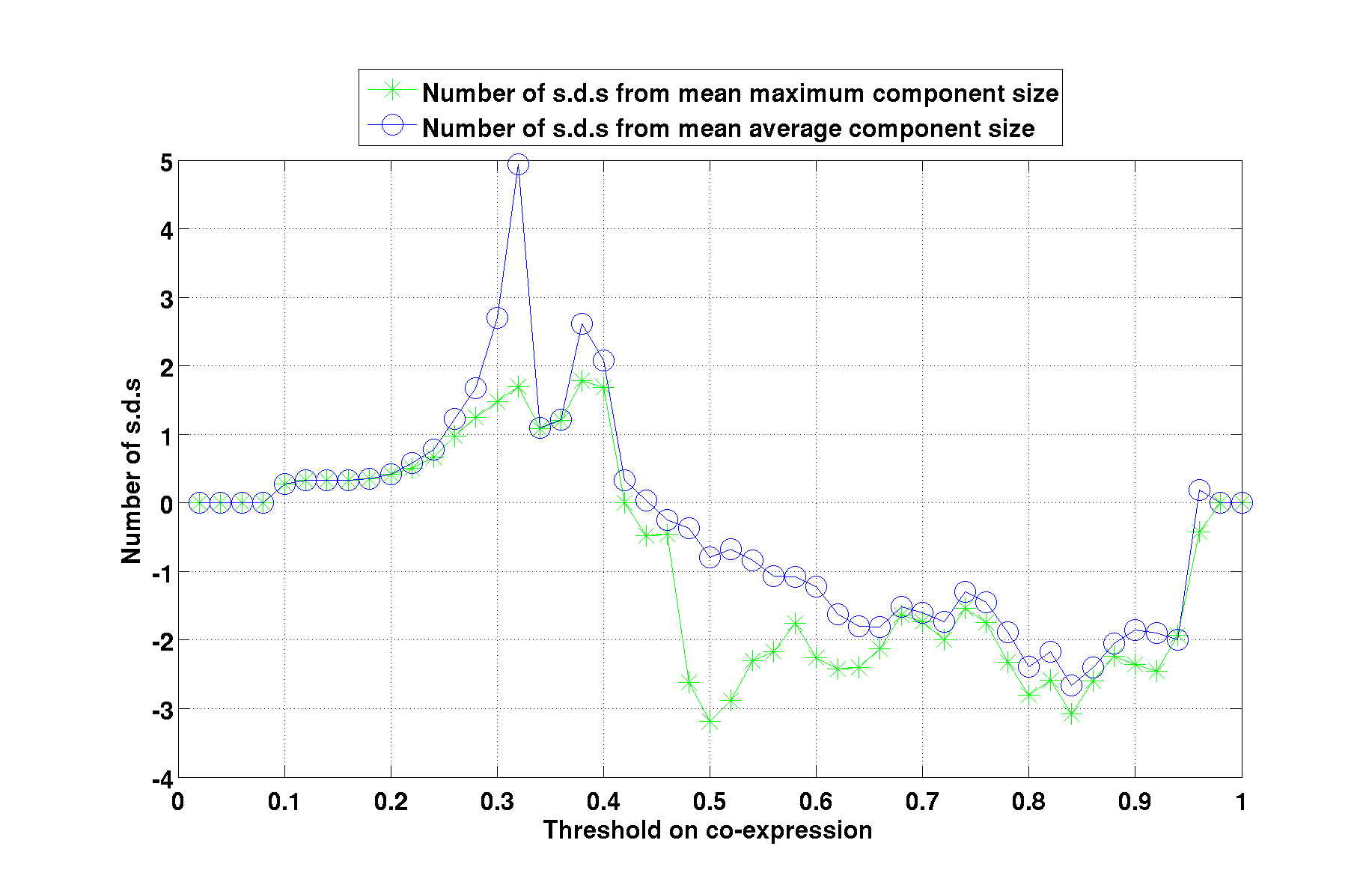}
\caption{{\bf{Monte Carlo analysis of the graph underlying the co-expression matrix of 288 genes 
 from the NicSNP database.}} Deviation of average and maximum size of connected components from mean (expressed in number of standard deviations at every value of the threshold).}
\label{coExpressionSet}
\end{figure}

\subsection{Neuroanatomical properties of connected components}
\subsubsection{Relation between fitting scores and co-expression}
If the co-expression of pairs of genes 
is defined as the cosine of the angle between their expression 
vectors in voxel space, as in Equation \ref{coExpressionFull}, the 
 fitting score of a gene $g$ to a region $\omega$ of the brain equals 
 the co-expression of gene $g$ and a (hypothetical) gene 
 whose expression profile would be proportional to
 the characteristic function of region $\omega$:
\begin{equation}
\phi_\omega( g ) = 1 - \frac{1}{2}\sum_{v\in\Omega}\left( E^{\mathrm{norm}}_g(v) - \chi_\omega( v ) \right)= 
       \sum_{v\in\Omega} E^{\mathrm{norm}}_g(v)  \chi_\omega( v ),
\label{fittingVersusCoExpression}
\end{equation}
where the second equality comes from the normalization of $E^{mathrm{norm}}_g$
 and $\chi_\omega$ in voxel space.

\subsubsection{Fitting scores of sums of gene-expression vectors}
At a given level of the threshold on co-expression,
a  set of genes is partitioned into connected components 
 induced by the graph underlying the thresholed matrix defined in 
Equation \label{threshEq}. For a set of genes of fixed size \lstinline$numGenes$, the function \lstinline$fitting_distribution_in_atlas.m$ estimates the distribution of fitting scores  of the sum of sets of genes of a given size, extracted from the Allen Atlas.\\

 In these functions, the genes are not weighted by coefficients to be optimized, they 
are simply summed over, so that the set of gene-expression
 vectors is different from the one explored in marker genes. The 
fitting score $\phi^{\mathrm{sum}}_\omega$ of the sum of genes depends
only on a list of $K$ distinct $K$ genes $\{ g_1,\dots,g_K \}$, and a region $\omega$ 
in the Allen Reference Atlas:
\begin{equation}
\phi_\omega(\{ g_1,\dots,g_K \} ) :=  \sum_{v\in\Omega} E^{\mathrm{norm}}_{\{ g_1,\dots,g_K \}}(v)\chi_\omega( v ),
\label{fittingScoreSum}
\end{equation}
where $E^{\mathrm{norm}}(\{ g_1,\dots,g_K \})$ is the sum 
 of gene-expression vectors of the genes $\{ g_1,\dots,g_K \}$, normalized 
 in the $L^2$ sense:
\begin{equation}
E^{\mathrm{norm}}_{{\{ g_1,\dots,g_K \}}}(v)= \frac{\sum_{i = 1}^K E( v, g_i)}{\sqrt{\sum_{v\in\Omega}
 \left(\sum_{j= 1}^K E( v, g_j) \right)^2}}.
\end{equation}

{\bf{Example \arabic{exampleCounter}. \addtocounter{exampleCounter}{1} Fitting score
 of a sum of genes in a given region.}} It can be computed using 
 the same functions as for the original data matrix. The sum of the relevant columns of the data matrix has to be substituted to the second argument of the function \lstinline$fitting_from_id.m$:\\

\begin{lstlisting}
colIndsForGenes = 1 : 10;
% the sum of these gene expressions (as a column vector )
sumOfDataForGenes = ( sum( ( E( :, colIndsForGenes ) )' )';
% consider the big12 annotation
identifierIndex = 5;
fittingScoresInAtlasBig12 = fitting_scores_region_by_gene( Ref, sumOfDataForGenes,...
 identifierIndex);
\end{lstlisting}

{\bf{Example \arabic{exampleCounter}. \addtocounter{exampleCounter}{1} Neuroanatomy 
 of nicotine-related genes in the \lstinline{big12} annotation.}}
\begin{lstlisting}
load( 'nicotineGenes.mat' );
indsSmall = nicotineGenes.nicotineIndicesInTop75;
coExpressionSmall = coExpressionFull( indsSmall, indsSmall );

% compute the connected components of the thresholded network
optionsComponents = struct( 'thresholdInit', 1, 'thresholdFinal', 0,...
 'thresholdStep', 0.02 );
coExpressionComponents = co_expression_components( coExpressionSmall,...
  optionsComponents );
% compare the fitting scores of the connected components 
% to the estimated distribution
% of fitting scores of sets of the same size
optionsAnatomy = struct( 'minimalComponentSize', 2, 'identifierIndex', 5,...
 'numDraws', 100 );
coExprComponentsAnatomy = co_expr_components_anatomy( Ref, E,...
  coExpressionComponents, indsSmall, optionsAnatomy );
\end{lstlisting}

One can search the results by $P$-value of fitting scores, 
 and/or size of connected components, and/or brain region:

{\bf{Example \arabic{exampleCounter}. \addtocounter{exampleCounter}{1} Neuroanatomy 
 of nicotine-related genes in the \lstinline{big12} annotation (continued).}}\\
\begin{lstlisting}
% show all the connected components, at any value of 
% the threshold, whose fitting score to any region is estimated to be in 
%
probaCrit = 0.001;
extractComponentByPValue = extract_component_by_p_value( Ref, E,...
 coExprComponentsAnatomy, probaCrit );
% illustrate the sums of these components
% and their anatomical properties
numComponentsExtracted = numel( extractByPvalue.threshold );
display( numComponentsExtracted )
if numComponentsExtracted > 0
  for kk = 1 : numComponentsExtracted
     figureComp = figure_for_component( Ref, E, extractByPvalue, kk );
     pause;
  end
end    
\end{lstlisting}

\chapter{Gene-based and cell-based expression data}
 The computational techniques  exposed in this chapter allow to reproduce the results
  of \cite{cellTypeBased,suppl1,suppl2,cerebellumCellTypesAutism,cellTypesRegionProc,cellTypesRegionPreprint} (see also \cite{KoCellTypes,TanFrenchPavlidis,JiCellTypes} 
   for analyses of the Allen Atlas data in terms of cell types).\\

\section{Cell-type-specific microarray data}
The file \lstinline$G_t_means.txt$ contains a matrix of microarray data.
 The rows correspond to the cell-type-specific samples coming from several studies
 and analyzed in \cite{OkatyComparison}. The columns correspond to genes (the
matrix is a type-by-gene matrix).
 The following code snippet (included in the file \lstinline$cell_type_start_up.m$)
  defines two matrices with the same numbers of columns, corresponding 
 to the intersection of the sets of genes in the coronal Allen Brain Atlas and in the 
 microarray data:
\begin{lstlisting}
% load the cell-type-specific data
C = dlmread( 'G_t_means.txt' );
% which columns of the data matrix C 
% correspond to genes that are also in E?
load( 'colsToUseInAllen.mat' );
load( 'colsToUseInTypes.mat' );
%construct the corresponding matrices
EUsed = E( :, colsToUseInAllen );
CUsed = C( :, colsToUseInTypes );
% Names of the cell types, in the same order as the rows of the matrix CUsed 
load( 'cellTypesDescription.mat' );
\end{lstlisting}

\section{Brain-wide correlation profiles}

The function \lstinline$cell_types_correls.m$ computes
 the voxel-by-type correlation matrix between the coronal Allen Brain Atlas
 and a set of cell types, as in equation \ref{cellTypesCorrelations}.
 
\begin{equation}
\mathrm{Corr}(v,t) = \frac{\sum_{g=1}^G( C(t,g) - \bar{C}(g))( E( v,g ) - \bar{E}(g) )}{\sqrt{\sum_{g=1}^G( C(t,g) - \bar{C}(g))^2}\sqrt{\sum_{g=1}^G ( E( v,g ) - \bar{E}(g) )^2}},
\label{cellTypesCorrelations}
\end{equation}
\begin{equation}
\bar{C}(g) = \frac{1}{T}\sum_{t=1}^T C(t,g),
\end{equation}
\begin{equation}
\bar{E}(g) = \frac{1}{V}\sum_{v=1}^V E(v,g).
\end{equation}

{\bf{Example \arabic{exampleCounter}. Brain-wide correlation profiles
 between cell-type-specific data and the Allen Atlas.\addtocounter{exampleCounter}{1}}}
 To compute the correlations for all the available cell types, the type indices 
specified in the variables span the whole range of row indices of the matrix 
 \lstinline$C$:\\
\begin{lstlisting}
% C is a type-by-gene matrix
numTypes = size( C, 1 );
typeIndices = 1 : numTypes;
cellTypesCorrelations = cell_types_correls( Ref, E, C,...
   colsToUseInAllen, colsToUseInTypes, typeIndices );
\end{lstlisting}

The results are saved in the file
\lstinline$cellTypesCorrelations.mat$.  It has $V=49,742$ rows and
$T=64$ columns (the matrix \lstinline$cellTypesCorrelations$ is a
voxel-by-type matrix, consistently 
with the definition of $\mathrm{Corr}(v,t)$ in Equation \ref{cellTypesCorrelations}), 
so that each column can be mapped to a volume
and visualized in exactly the same way as a column of the data matrix
\lstinline$E$:\\ 

{\bf{Example \arabic{exampleCounter}. Visualization of brain-wide
 correlation profiles.\addtocounter{exampleCounter}{1}}} Let us
plot maximal-intensity projections of the correlations between the
20-th row of $C$ and all the rows of the Allen Atlas.\\
\begin{lstlisting}
% load the correlations, unless of course they have just been
% computed using the previous example
load( 'cellTypesCorrelations.mat' );
% take a colum of the correlation matrix,
% i.e. focus on a specific cell type
typeIndex = 20;
% describe the cell type specified by typeIndex
display( cellTypesDescription( typeIndex ) )
correlsTypeToAtlas = cellTypesCorrelations( :, typeIndex );
% map the column to a three-dimensional grid
correlsTypeToAtlasVol = make_volume_from_labels( correlsTypeToAtlas,...
                 brainFilter );
plot_intensity_projections( correlsTypeToAtlasVol );
\end{lstlisting}

\section{Brain-wide density estimates of cell types}

To decompose the gene-expression at every voxel in the Allen Atlas
into its cell-type-specific components, let us introduce the positive
quantity $\rho_t( v )$ denoting the contribution of cell-type $t$ at
voxel $v$, and propose the following linear model:
\begin{equation}
E( v,g) = \sum_{ t = 1 }^T \rho_t( v )C_t(g) + {\mathrm{Residual(v,g)}}.
\label{linearModel}
\end{equation}
Both sides are estimators of the number of mRNAs for gene $g$ at voxel $v$.  
The residual term in Equation \ref{linearModel} reflects the 
 fact that $T=64$ cell types are not enough to sample the whole diversity of cell types 
 in the mouse brain, as well as noise in the measurements, reproducibility
issues, the non-linearity of the relations between numbers of mRNAs,
 expression energies and microarray data.\\

To find the best fit of the model, we have to minimize the residual term
by solving the following problem :
\begin{equation}
\left(\rho_t( v )\right)_{1\leq t \leq T, 1\leq v \leq V} =
     {\mathrm{argmin}}_{\phi\in {\mathbf{R}}_+(T,V)}\mathcal{
       E}_{E,C}(\phi ),
\label{fittingProblem}
\end{equation}
where 
\begin{equation}
\mathcal{E}_{E,C}( \phi ) = \sum_{v=1}^V\left( \sum_{g=1}^G \left(
E(v,g) - \sum_{t=1}^T\phi(t,v)C_t(g)\right)^2 \right).
\label{errorTerm}
\end{equation}
As the terms of the sum over voxels that involve a fixed voxel $v$
involve only terms of the test function $\phi$ with the same voxel index,
 and are positive, the
problem \ref{fittingProblem} can be solved voxel by voxel. For each voxel we have to
minimize a quadratic function of a vector with $T$ positive
components.
\begin{equation}
\forall v \in [1..V], (\rho_t( v ))_{1\leq t \leq T}=
        {\mathrm{argmin}}_{\nu\in {\mathbf{R}}_+^T}\sum_{g=1}^G \left(
        E(v,g) - \sum_{t=1}^T\nu(t) C_t(g)\right)^2.
\label{voxelByVoxel}
\end{equation}
For each cell type $t$, the coefficients $(\rho_t( v ))_{1\leq v \leq V}$ yield a brain-wide
fitting profile between this cell type and the Allen Atlas.\\

The function \lstinline$fit_voxels_to_types_picked.m$ solves the quadratic optimization problems
 of Equation \ref{voxelByVoxel}, one per voxel, for a list of cell-type indices, and returns the result in a voxel-by-type 
format.\\

{\bf{Example \arabic{exampleCounter}.\def\fitExLabel{\value{exampleCounter}}
 \addtocounter{exampleCounter}{1}}} Consider the full
  cell-type-specific dataset with \lstinline$numTypes$ cell types, and 
 estimate the brain-wide density of each of these cell types:
\begin{lstlisting}
numTypes = size( C, 1 );
typeIndicesPicked = 1 : numTypes;
fitVoxelsToTypesFull = fit_voxels_to_types_picked( Ref, E, C,...
 colsToUseInAllen, colsToUseInTypes, typeIndicesPicked );
\end{lstlisting}
The result of the computation in the above code snippet 
 are saved in the file \lstinline$fitVoxelsToTypes.mat$. Note that the results 
depends on the set of types specified by \lstinline$typeIndicesPicked$, since the corresponding 
 columns of $C$ play the role of a basis on which the gene-expression vector at each voxel is
 decomposed under positive constraints. \\

\section{Cell-type-based computational neuroanatomy}
As the matrix of cell-type-specific microarray data is presented 
 in the type-by-gene format, a given cell type in this dataset corresponds
 to a row index $t$ in  $[ 1, T ]$. The corresponding column of any voxel-by-type matrix
 (such as correlations ${\mathrm{Corr}}$, Equation \ref{cellTypesCorrelations},
 or densities $\rho$, Equation \ref{linearModel})
can be mapped to volumes using the function \lstinline$make_volume_from_labels.m$.
 Projections of these volume can be visualized 
 using the function \lstinline$plot_intensity_projections.m$. One can
   supplement these projections with 
 sections of these volumes. The sections can be chosen computationally as 
 follows by ranking the brain regiions in the Allen Reference Atlas using the
 brain-wide correlation or density profiles.\\

{\bf{Ranking of regions by correlations with a cell type.}} Given a
three-dimensional grid containing brain-wide correlations between a
given cell type $t$ and the Allen Atlas (as defined in Equation \ref{cellTypesCorrelations}), one can rank the regions in a
given annotation of the Allen Reference Atlas by computing the average
correlation across in each of the regions:
 \begin{equation}
{\overline{\mathrm{Corr}}}( r, t ) = \frac{1}{|V_r|}\sum_{v \in V_r}{\mathrm{Corr}}(v,t).
\label{correlRegion}
\end{equation}
For each cell type $t$, each region $r$ in the ARA is ranked
according to its average correlation. Call its rank $\chi( r )$. In particular,
the correlation profile yields a top region $\rho^\chi(t)$ in the ARA, for which 
$\chi( \rho^\chi(t) ) = 1$. This is the region whose voxels
 are most highly correlated to cell type $t$ on average.\\

{\bf{Ranking of regions by estimated density of a cell type.}}
 Consider a system of annotation of the ARA, taken from Table \ref{annotationSystems}
 \lstinline$Big12$ is the default option in the Matlab implementation).
 For each cell-type index $t$ and each region $V_r$ (where $r$ is an 
 integer index taking values in $[1..R]$, where $R$ is the number of regions
 in the chosen system of annotation), we can compute the 
 contribution of the voxels of the region to the brain-wide density profile 
 of  cell type $t$:
\begin{equation}
{\overline{\rho}}( r, t ) = \frac{1}{\sum_{w \in {\mathrm{Brain}}}\rho_t( w )}
                                  \sum_{v \in V_r}\rho_t( v ).
\label{fittingRegion} 
\end{equation}
Given a cell type $t$, the regions in the ARA is ranked according to its average 
correlation. Call its rank $\phi( r )$. In particular,
the correlation profile yields a top region $\rho^\phi(t)$ in the ARA, for which 
$\phi(\rho^\phi(t)  ) = 1$. This is the region whose voxels bring the 
 largest total contribution to the vector $\rho_t$.\\ 

These two rankings are implemented in the function \lstinline$classify_pattern.m$ (see example
 below). The values of the fraction of total density and average correlation
 for a given cell type can be plotted (in the \lstinline$big12$ version of the 
 ARA), using\footnote{The labels used in these plots
 can be decoded by comparing the fields \lstinline$Ref.Coronal.Annotations.labels{5}$
  and \lstinline$Ref.Coronal.Annotations.symbols{5}$. This yields
 the following correspondences: \lstinline$Basic cell groups and regions = Brain,
 Cerebral cortex = CTX,$\\
\lstinline$ Olfactory areas = OLF, Hippocampal region = HIP, Retrohippocampal region = RHP, Striatum = STR,$\\
\lstinline$
 Pallidum = PAL, Thalamus = TH, Hypothalamus = HY,  Midbrain = MB,  Pons = P,$\\
\lstinline$ Medulla = MY, Cerebellum = CB$.} the functions \lstinline$plot_correlations_for_big12.m$
 and \lstinline$plot_densities_for_big12.m$.\\

\begin{figure}
\centering
\includegraphics[width=\figWidth,keepaspectratio]{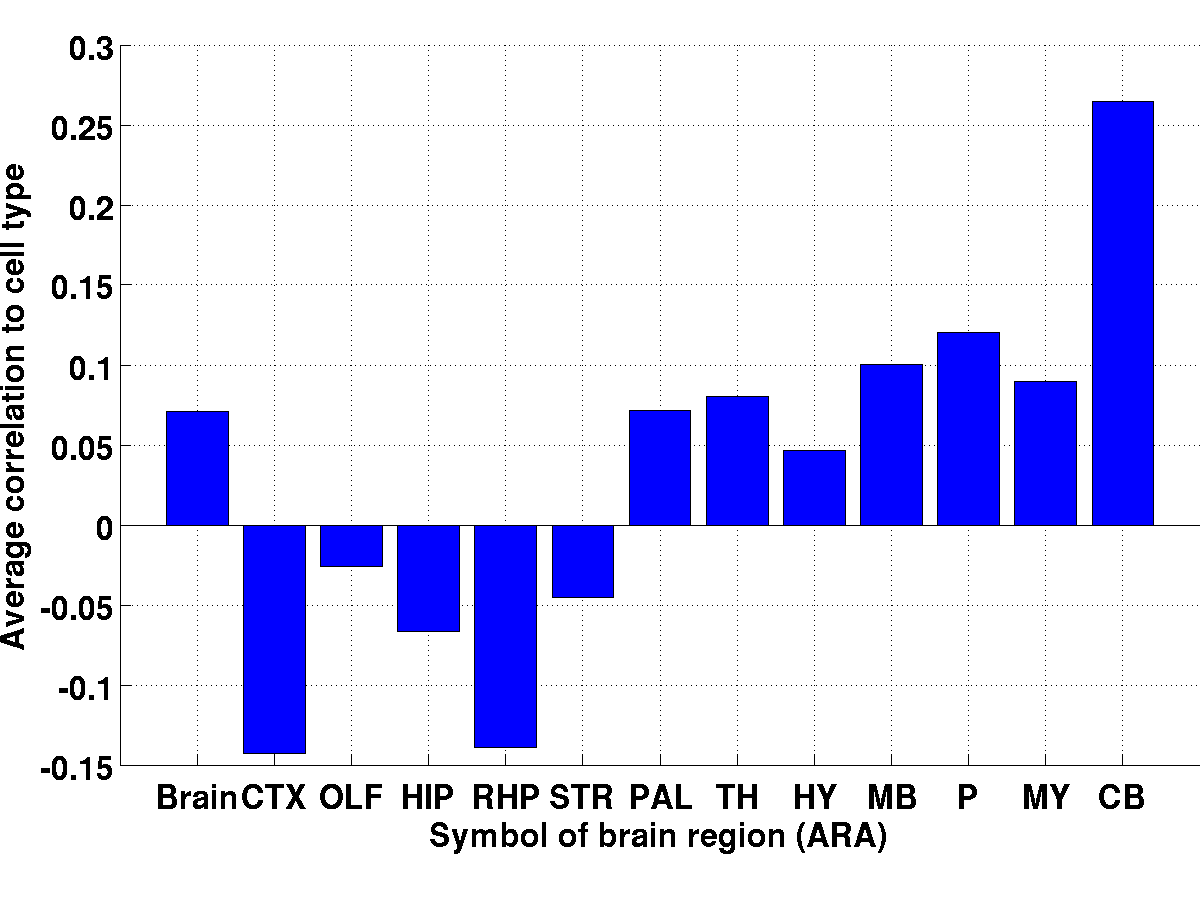}
\caption{Average correlations between the voxels of the coronal  Allen Atlas (G=3,041 genes) and granule cells (index 20 in the cell-type-specific dataset), in the regions of the \lstinline$big12$ annotation of the ARA.}
\label{avgCorrelationGranuleCells}
\end{figure}

\begin{figure}
\centering
\includegraphics[width=\figWidth,keepaspectratio]{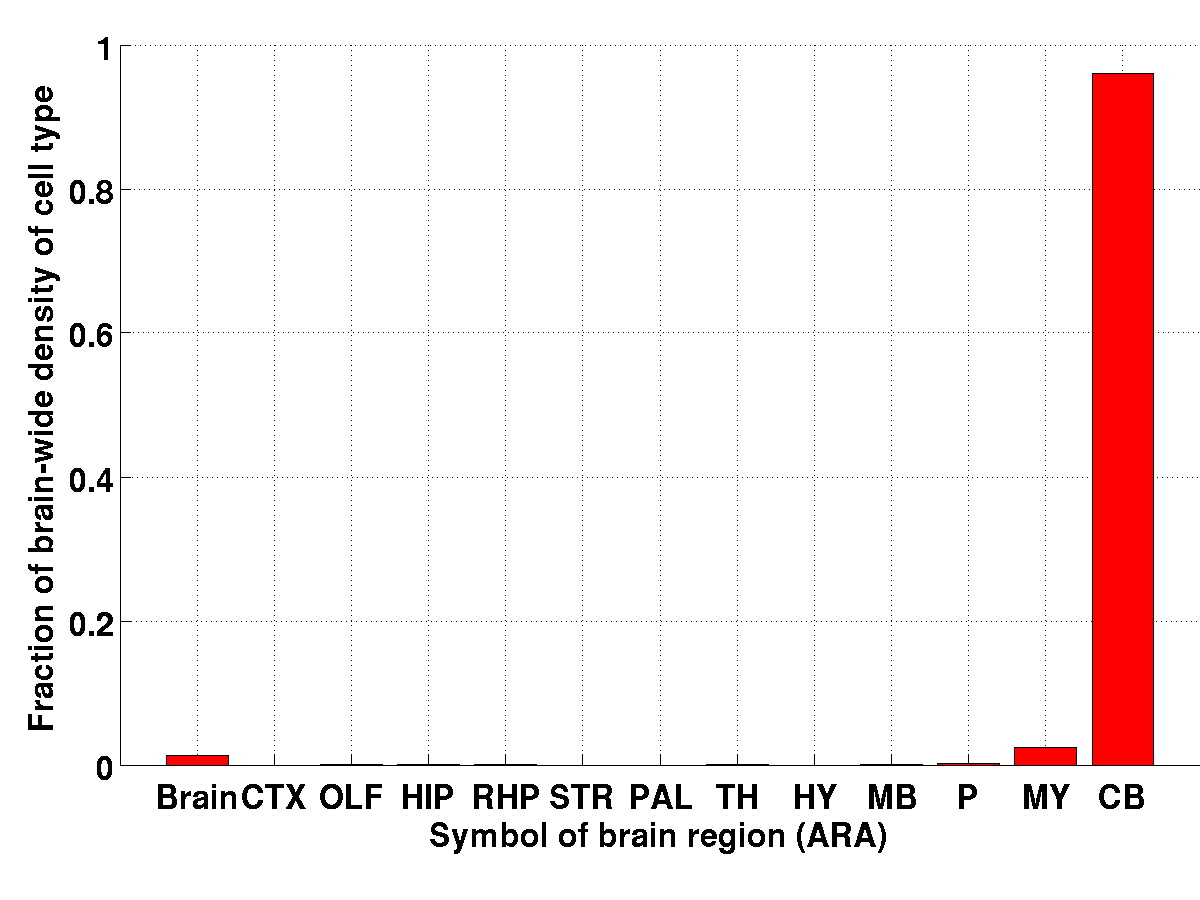}
\caption{Fraction of the density of granule cells (index 20 in the cell-type-specific dataset)
 conributed by the regions in the \lstinline$big12$ annotation of the ARA.}
\label{fracDensityGranuleCells}
\end{figure}

{\bf{Example \arabic{exampleCounter}.\addtocounter{exampleCounter}{1}}} 
Rank the regions of the \lstinline$big12$ annotations by average
correlation and fraction of the density of cell types, and display 
 the results for granule cells. The code snippet 
 below should reproduce Figures \ref{avgCorrelationGranuleCells} and 
\ref{fracDensityGranuleCells}.
\begin{lstlisting}
% Rank brain regions (in the big12 annotation of the ARA)
% by their correlation and fitting profiles to each cell type
classifyOptions = struct( 'identifierIndex', 5 );
classifyBig12 = classify_pattern( Ref, cellTypesCorrelations,...,
                  fitVoxelsToTypes, classifyOptions );
% the values of average correlation and fraction of density
regionByTypeAvgCorrelation = classifyBig12.regionByTypeAverageCorrelations;
regionByTypeFracDensity = classifyBig12.regionByTypeFractionFittings;
% the ranks of the regions, in the same region-by-type format
rankRegionsByCorrelation = classifyBig12.rankRegionsAverageCorrelations;
rankRegionsByDensity = classifyBig12.rankRegionsFractionFittings;
% Focus on the granule cells, index 20;
typeIndex = 20;
display( cellTypesDescription( typeIndex ) );
granuleCellsAvgCorrelation = regionByTypeAvgCorrelation( :, typeIndex );
granuleCellsCorrelationRanks = rankRegionsByCorrelation( :, typeIndex );
granuleCellsFracDensity = regionByTypeFracDensity( :, typeIndex );
granuleCellsRankRegionsByDensity = rankRegionsByDensity( :, typeIndex );
identifierIndex = classifyOptions.identifierIndex;
labels = ann.labels{ identifierIndex };
granuleCellsRankedRegionsCorr = labels( granuleCellsCorrelationRanks );
% regions in the big12 annotations, ranked by correlation
display( granuleCellsRankedRegionsCorr );
granuleCellsRankedRegionDensity = labels( granuleCellsRankRegionsByDensity );
% regions in the big12 annotations, ranked by correlationdensity
display( granuleCellsRankedRegionDensity );
%confirm these rankings by plotting the underlying values
plot_correlations_for_big12( Ref, granuleCellsAvgCorrelation );
pause( 2 );
hold off;
plot_densities_for_big12( Ref, granuleCellsFracDensity );
hold off;
\end{lstlisting}

Given a region in the ARA (our best guess according to one of the above criteria),
 we need to pick a section that intersects this region. The
 function \lstinline$cell_type_vol_prepare.m$ implements 
the choice of the section. It chooses 
 the section (which can be sagittal, coronal or axial, specified by 
 the field \lstinline$options.sectionStyle$ of the options)
 that intersect the desired region (specified by the 
 fields \lstinline$options.identifierIndex$ and 
\lstinline$options.regionIndexForSection$) along the 
 largest number of voxels, unless the field 
  \lstinline$options.customIndex$ equals 1. Then it 
 works out the section (of the required style) 
 at a position specified by \lstinline$option.desiredIndex$.
  This function is used by the function \lstinline$figure_for_types.m$
 to produce Figures like \ref{figureTypeCorrels47} and \ref{figureTypeDensity47}.


\section{Visualization of correlation and density profiles of cell types}
\begin{figure}
\centering
\includegraphics[width=\figWidth,keepaspectratio]{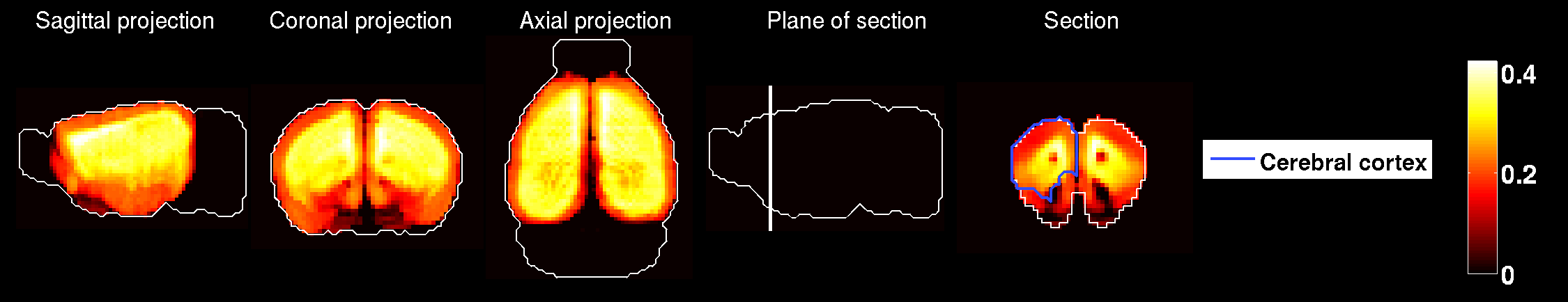}
\caption{{\bf{Correlations between the voxels of the coronal  Allen Atlas (G=3,041 genes) and pyramidal cells (index 47 in the cell-type-specific dataset).}} Maximal-intensity projection and a section through the region in the ARA that maximizes the average correlation.}
\label{figureTypeCorrels47}
\end{figure}

{\bf{Example \arabic{exampleCounter}. \addtocounter{exampleCounter}{1}}} Consider the pyramidal neurons, index 47 in the auxiliary 
dataset. The following code snippet computes the region in \lstinline$big12$
annotation of the ARA with the highest average correlation (as in Equation
 \ref{correlRegion}), and chooses the section of the brain through 
 this section that maximizes the (section-wide) average correlation. It displays it 
 next to the maximal-intensity projections of the correlation profiles, together 
with a graphical definition of the plane that was used to obtain the section.
 The snippet should reproduce Figure \ref{figureTypeCorrels47} 
and save it as \lstinline$figureTypeCorrels47.png$.
\begin{lstlisting}
typeIndex = 47;
display( cellTypesDescription( typeIndex ) )
correlsTypeToAtlas = cellTypesCorrelations( :, typeIndex );
correlsTypeToAtlasVol = make_volume_from_labels( correlsTypeToAtlas,...
 brainFilter );
optionsFig = struct( 'identifierIndex', 5, 'sectionStyle', 'coronal' );
figureForTypeCorrelations = figure_for_type_correlations( Ref,...
  correlsTypeToAtlasVol, optionsFig );
\end{lstlisting}

\begin{figure}
\centering
\includegraphics[width=\figWidth,keepaspectratio]{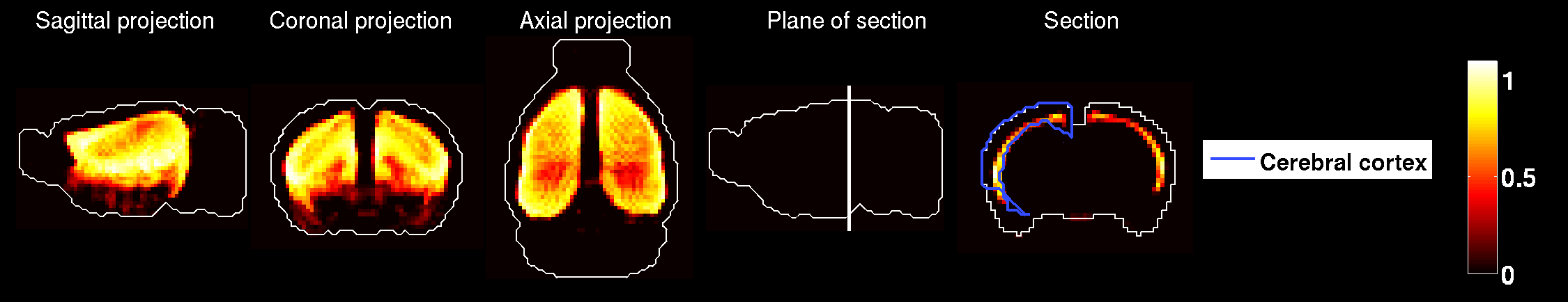}
\caption{{\bf{Brain-wide density estimate of 
 pyramidal cells (index 47 in the cell-type-specific dataset).}} Maximal-intensity projection and a section through the region in the ARA that contains the largest fraction of total density.}
\label{figureTypeDensity47}
\end{figure}

{\bf{Example \arabic{exampleCounter}. \addtocounter{exampleCounter}{1}}} Consider again the pyramidal neurons, index 47 in the auxiliary 
dataset. The following code snippet allows to visualize a section 
 of their estimated brain-wide density, through the region
of the ARA that contains the largest fraction of the total density,
as per Equation \ref{fittingRegion}. It should reproduce Figure 
\ref{figureTypeDensity47} 
and save it as \lstinline$figureTypeDensity47.png$.

\begin{lstlisting}
typeIndex = 47;
densityOfType = fitVoxelsToTypes( :, typeIndex );
densityOfTypeVol = make_volume_from_labels( densityOfType, brainFilter );
figureForTypeDensity = figure_for_type_density( Ref,  densityOfTypeVol,...
 optionsFig );
set( gcf, 'PaperPositionMode', 'auto' );
saveas( gcf, 'figureTypeDensity47.png', 'png' );
\end{lstlisting}

\chapter{Clustering}
\section{Kullback--Leibler distance from gene-expression to a given probability distribution}
One can use the Kullback--Leibler (KL) divergence to compare the brain-wide expression of
a gene to a given probability distribution over the brain.\\

 This probability distribution can be for instance
 the normalized average $E_{\mathrm{full}}^{\mathrm{avg}}$ across all genes in the coronal atlas for instance

\begin{equation}
{\mathrm{KL}}^{\mathrm{avg}}(g) : = \sum_{v=1}^V \tilde{E}_g(v)
   \log\left( \frac{\tilde{E}_g(v)}{\tilde{E}_{\mathrm{full}}^{\mathrm{avg}}(v)}    \right),
\end{equation} 
where $\tilde{E}_{\mathrm{full}}^{\mathrm{avg}}$ and $\tilde{E}_g$
are probability densities over the brain obtained by normalization:
 \begin{equation}
 \tilde{E}_g(v) = \frac{E(v,g)}{ \sum_{u=1}^V E(u,g ) },\;\;\;\;
\tilde{E}_{\mathrm{full}}^{\mathrm{avg}}(v) = \frac{E_{\mathrm{full}}^{\mathrm{avg}}(v)  }{ \sum_{u=1}^V E_{\mathrm{full}}^{\mathrm{avg}}(u)}.
\end{equation}

\section{Construction of a bipartite graph from a voxel-by-gene matrix}

The Allen Atlas can be mapped to a \emph{weighted bipartite graph} in the following way:
\begin{itemize}
\item the first set of vertices consists of voxels, numbered from $1$ to $V$,
\item the second set of vertices consists of genes, numbered from $1$ to $G$,
\item each of the  edges connects one voxel, say $v$ to one gene, say $g$ and has a weight given by the expression energy of $E(v,g)$ of the gene at the voxel.\\
\end{itemize}
We looked for partitions of this weighted bipartite graph into
subgraphs such that the weights of the internal edges of the subgraphs
are strong compared to the weights of the edges between the
subgraphs. This is the isoperimetric problem addressed by the
algorithm of \cite{bi-clustering} (the graph need not be bipartite to
apply this algorithm, but since we started with a bipartite graph,
each of the subgraphs, or biclusters, returned by the algorithm, is
bipartite, and therefore corresponds to a set of voxels and a set of
genes).\\

Given a weighted graph, the algorithm cuts some of the links, thus partitioning the graph 
into a subset $S$ and its complementary $\bar{S}$,  
such that the sum of weights in the set of cut edges is minimized relative to the 
total weight of internal edges in $S$. The sum of weights in the set of cut edges is analogous to a boundary term,
while the total weight of  internal edges is analogous to a volume term. In that sense the problem
is an isoperimetric optimization problem, and the optimal set $S$ minimizes the {\emph{isoperimetric ratio}} $\rho$ over
 all the possible subgraphs:
\begin{equation}
S = {\mathrm{argmin}}_{{\mathrm{Vol}}(s) \leq {\mathrm{Vol}}(\bar{s}) } \,\rho( s ),
\end{equation}
\begin{equation}
\rho( s ) := \frac{|\partial s|}{\mathrm{Vol}(s)},
\end{equation}
\begin{equation}
|\partial s| = \sum_{i\in s, j \in \bar{ s}} W_{ij},
\end{equation}
\begin{equation}
{\mathrm{Vol}}(s) = \sum_{ i\in s, j\in s} W_{ij},
\end{equation}
 where the quantity $W_{ij}$ is the weight of the link between vertex $i$ and vertex $j$. 
Once $S$ has been worked out, the algorithm can be applied separately to $S$ and its complementary $\bar{S}$. This 
recursive application goes on until the isoperimetric ratio reaches a stopping ratio, representing the 
highest allowed isoperimetric ratio. This value is a parameter of the 
algorithm. Rising it results in a higher number of clusters, as it rises the number of acceptable cuts. 
 The implementation of the recursive partition of the bipartite graph
 in the present toolbox is due to Grady and Schwartz \cite{bi-clustering}.\\

{\bf{Example \arabic{exampleCounter}. Partition the left hemisphere of the 
 brain according to the expression of a given set of genes,
 chosen for their high divergence from a uniform
 expression.\addtocounter{exampleCounter}{1}}}
The code snippet below should reproduce 
 Figures \ref{geneSumBicluster1,geneSumBicluster2,geneSumBicluster3}, 
among other things.
\begin{lstlisting}
indsBrain = 1 : numel( brainFilter );
ann = cor.Annotations;
% consider the row indices in the data matrix constructed from
% voxels in the left hemisphere ('big12' annotation)
leftFilter = get_voxel_filter( cor, ann.filter( 5 ) );
indsBrain =  make_volume_from_labels( indsBrain, brainFilter ); 
indsLeft = indsBrain( leftFilter );
% normalise the sum of each column of the
% data matrix so that each corresponds corresponds to a 
% probability  density in voxel space 
ENorm = normalise_integral( E ); 
%  rank the genes by descending order of localization 
klDivergence = KL_divergence_from_uniform( ENorm ); 
[ vals, inds ] = sort( klDivergence, 'descend' ); 
plot_intensity_projections( make_volume_from_labels( E( :, inds( 1 ) ), brainFilter ) );
% keep the numGenesKept most localized genes 
numGenesKept = 150;
numGenesTot = size( E, 2 ); 
indsTaken = inds( 1 : numGenesKept );
voxelByGeneMatrix = E( indsLeft, indsTaken ); 
% for speed and memory, keep the most intense
% voxels construct a bipartite graph (the thresholdRank voxels
% with highest total energy kept)
voxelIntensities = sum( voxelByGeneMatrix, 2 );
[ valsInd, indsInt ] = sort( voxelIntensities );
thresholdRank = 6351; 
voxelIndsKept = indsInt( 1 : thresholdRank )
adjacencyMatrix = biclustering_graph_voxel_by_gene( voxelByGeneMatrix, voxelIndsKept );
% consistency check for the size of the adjacency matrix
numInGraph = size( adjacencyMatrix, 1 );
numVoxKept = numel( voxelIndsKept ); 
numVoxKeptComputed = numInGraph - numGenesKept; 
if numVoxKept ~= numVoxKeptComputed 
  display( [ numVoxKept , numVoxKeptComputed ] );
  error( 'size of adjacency matrix inconsistent with data' );
end 
% choose a range of stopping criteria (numbers between 0 and 1:
% maximum fraction of the weight of links that allowed to be cut
% before the algorithm terminates
stoppingCrits = 0.2 : 0.02 : 0.3;;
for ss = 1 : numel( stoppingCrits ) 
  [ voxels, genes, cluster ] = bicluster_graph( adjMat, 'iso', stoppingCrits( ss ),...
  numVoxKeptComputed ); 
  myClusters.voxels{ ss } = voxels;
  myClusters.genes{ ss } = genes;
  myClusters.cluster{ ss } = cluster;
  myClusters.stoppingCrit{ ss } = stoppingCrits( ss ); 
  % the number of clusters should be a growing function 
  % of the stopping criterion
  numClust( ss ) = numel( cluster );
end
% visualize the sums of gene-expressions 
% in each of the clusters
for ss = 1 : numel( stoppingCrits ) 
  stoppingCriterionUsed = stoppingCrits( ss );
  display( stoppingCriterionUsed )
  cluster = myClusters.cluster{ ss };
  for cc = 1 : numel( cluster )
     geneIndsInTaken = cluster( cc ).geneIdx;
     indsInAtlas = indsTaken( geneIndsInTaken );
     geneSum = sum( E( :, indsInAtlas ), 2 ); 
     plot_intensity_projections( make_volume_from_labels( geneSum,...
        brainFilter ) ); 
  end
  pause;
  hold  off;
  close all; 
end 
\end{lstlisting}

\begin{figure}
\centering
\includegraphics[width=\figWidth,keepaspectratio]{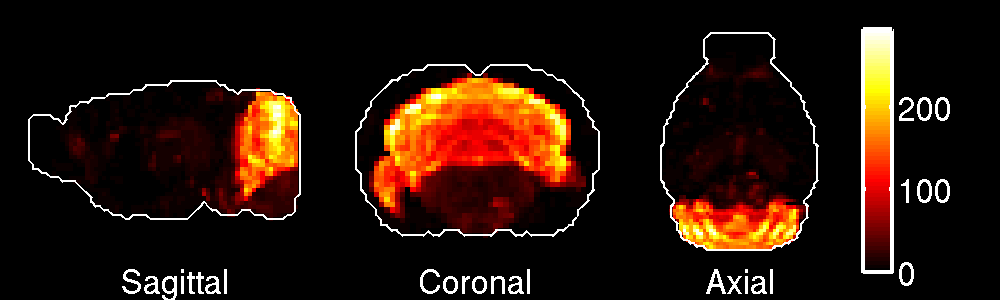}
\caption{Maximal-intensity projections of the brain-wide sum of gene-expression
 for the genes in the first bicluster returned by the Example \arabic{exampleCounter}.}
\label{geneSumBicluster1}
\end{figure}
\begin{figure}
\centering
\includegraphics[width=\figWidth,keepaspectratio]{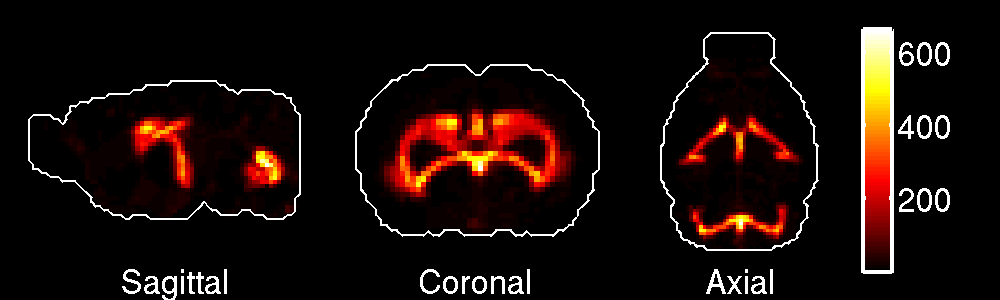}
\caption{Maximal-intensity projections of the brain-wide sum of gene-expression
 for the genes in the second bicluster returned by the Example \arabic{exampleCounter}.}
\label{geneSumBicluster2}
\end{figure}
\begin{figure}
\centering
\includegraphics[width=\figWidth,keepaspectratio]{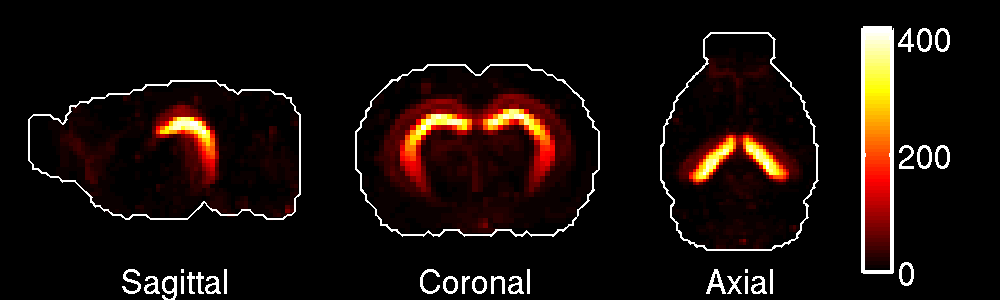}
\caption{Maximal-intensity projections of the brain-wide sum of gene-expression
 for the genes in the third bicluster returned by the Example \arabic{exampleCounter}.}
\label{geneSumBicluster3}
\end{figure}
\begin{figure}
\centering
\includegraphics[width=\figWidth,keepaspectratio]{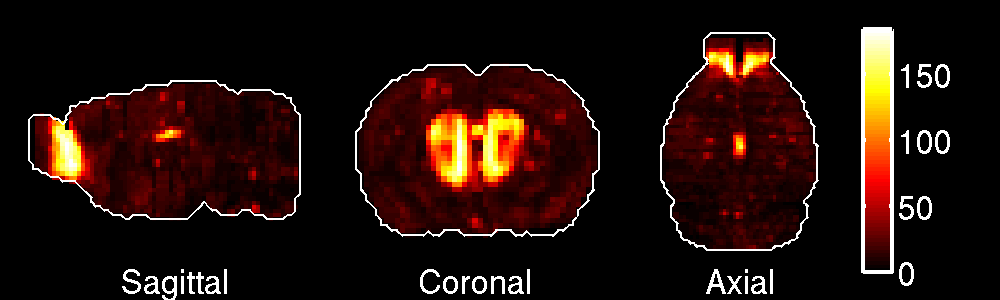}
\caption{Maximal-intensity projections of the brain-wide sum of gene-expression
 for the genes in the first bicluster returned by the Example \arabic{exampleCounter}.}
\label{geneSumBicluster1}
\end{figure}
\begin{figure}
\centering
\includegraphics[width=\figWidth,keepaspectratio]{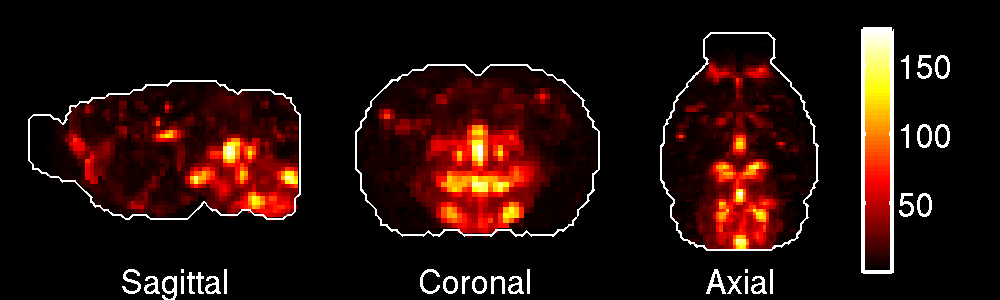}
\caption{Maximal-intensity projections of the brain-wide sum of gene-expression
 for the genes in the second bicluster returned by the Example \arabic{exampleCounter}.}
\label{geneSumBicluster2}
\end{figure}
\begin{figure}
\centering
\includegraphics[width=\figWidth,keepaspectratio]{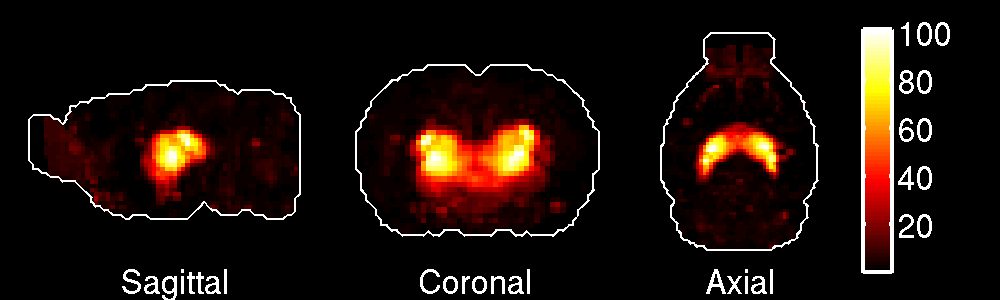}
\caption{Maximal-intensity projections of the brain-wide sum of gene-expression
 for the genes in the third bicluster returned by the Example \arabic{exampleCounter}.}
\label{geneSumBicluster3}
\end{figure}

\chapter{List of files}
\section{Preparation of data}
\lstinline$mouse_start_up.m$\\
\lstinline$cell_type_start_up.m$\\
\section{Atlas navigation}
\lstinline$get_annotation.m$\\
\lstinline$get_genes.m$\\
\lstinline$get_voxel_filter.m$\\
\lstinline$make_volume_from_labels.m$\\
\lstinline$plot_intensity_projections.m$\\
\lstinline$flip_through_sections.m$\\
\lstinline$outline_section_with_atlas_picked.m$\\
\lstinline$annotation_big12_to_fine$\\
\lstinline$annotation_fine_to_big12.m$\\
\section{Marker genes}
\lstinline$localization_from_id.m$\\
\lstinline$localization_scores_region_by_gene.m$\\
\lstinline$fitting_from_id.m$\\
\lstinline$fitting_scores_region_by_gene.m$\\
\lstinline$find_find_lambdamax_l1_ls_nonneg.m$ (author K. Koh)\\
\lstinline$l1_ls.m$ (author K. Koh)\\
\lstinline$l1_ls_nonneg.m$ (author K. Koh)\\
\section{Co-expression analysis}
\lstinline$co_expression_matrix.m$\\
\lstinline$upper_diagonal_coeffs.m$\\
\lstinline$cumul_co_expr.m$\\
\lstinline$cumul_co_expr_plot.m$\\
\lstinline$co_expression_components_bootstrap.m$\\
\lstinline$co_expression_components_bootstrap_plot.m$\\
\section{Clustering}
\lstinline$KL_divergence_from_uniform.m$\\
\lstinline$biclustering_graph_voxel_by_gene.m$
\lstinline$bicluster_graph.m$\\
\lstinline$partitiongraph.m$ (authors: A. Pothen, H. Simon and K.-P. Liou \cite{partitionGraph})\\
\lstinline$recursivepartition.m$ (author: L. Grady)\\
\lstinline$isosolve.m$ (author: L. Grady)\\
\section{Cell-type specific analysis}
\lstinline$cell_type_start_up.m$\\
\lstinline$cell_types_correls.m$\\
\lstinline$fit_voxels_to_types_picked.m$\\
\lstinline$cell_type_vol_prepare.m$\\
\lstinline$classify_pattern.m$\\
\lstinline$figure_for_types.m$\\
\lstinline$figure_for_type_correlations.m$\\
\lstinline$figure_for_type_correlationsdensity.m$\\
\lstinline$plot_correlations_for_big12.m$\\
\lstinline$plot_densities_for_big12.m$\\
\section{Miscellaneaous operations on matrices}
\lstinline$normalise_integral.m.m$\\
\lstinline$normalise_integral_L2.m$\\
\section{Data files}
\subsection{Allen Brain Atlas, Allen Reference Atlas}
\lstinline$refDataStruct.mat$\\
\lstinline$ExpEnergy.mat$\\
\lstinline$ExpEnergytop75percent.mat$\\

\subsection{Cell-type specific analysis}
\lstinline$G_t_means.txt$\\
\lstinline$colsToUseInAllen.mat$\\
\lstinline$colsToUseInTypes.mat$\\
\lstinline$cellTypesDescription.mat$\\

\section{Saved results}
\subsection{Marker genes}
\lstinline$quadTotIdentifier5.mat$\\
\lstinline$quadRegIdentifier5Region1.mat$\\ 
\lstinline$quadRegIdentifier5Region2.mat$\\ 
\lstinline$quadRegIdentifier5Region3.mat$\\
\lstinline$quadRegIdentifier5Region4.mat$\\
\lstinline$quadRegIdentifier5Region5.mat$\\
\lstinline$quadRegIdentifier5Region6.mat$\\
\lstinline$quadRegIdentifier5Region8.mat$\\
\lstinline$quadRegIdentifier5Region9.mat$\\
\lstinline$quadRegIdentifier5Region10.mat$  \\
\lstinline$quadRegIdentifier5Region11.mat$  \\
\lstinline$quadRegIdentifier5Region12.mat$ \\
\lstinline$quadRegIdentifier5Region13.mat$ \\

\subsection{Cell-type specific analysis}
\lstinline$cellTypesCorrelations.mat$\\
\lstinline$fitVoxelsToTypes.mat$\\

\subsection{Playground}
\lstinline$tool_box_manual_illustration.m$\\

\end{document}